# Differentially Categorized Structural Connectome Hubs are Involved in Differential Microstructural Basis and Functional Implications and Contribute to Individual Identification

Xindi Wang[1,2], Qixiang Lin[1,2], Mingrui Xia[1,2,*], Yong He[1,2,*]

[1] National Key Laboratory of Cognitive Neuroscience and Learning, Beijing Normal University, Beijing, 100875 China

[2] IDG/McGovern Institute for Brain Research, Beijing Normal University, Beijing, 100875 China

Running Title: Differentially Categorized Structural Connectome Hubs are Involved in Differential Microstructural Basis and Functional Implications and Contribute to Individual Identification

* Corresponding Authors:

Mingrui Xia, Ph.D.,
National Key Laboratory of Cognitive Neuroscience and Learning, IDG/McGovern Institute for Brain Research, Beijing Normal University, Beijing, 100875, China. E-mail: mxia@bnu.edu.cn,

Yong He, Ph.D.,
National Key Laboratory of Cognitive Neuroscience and Learning, IDG/McGovern Institute for Brain Research, Beijing Normal University, Beijing, 100875, China. E-mail: yong.he@bnu.edu.cn

# Abstract

Human brain structural networks contain sets of centrally embedded hub regions that enable efficient information communication. However, it remains largely unknown about categories of structural brain hubs and their microstructural, functional and cognitive characteristics as well as contributions to individual identification. Here, we employed three multi-modal imaging data sets with structural MRI, diffusion MRI and resting-state functional MRI to construct individual structural brain networks, identify brain hubs based on eight commonly used graph-nodal metrics, and perform comprehensive validation analysis. We found three categories of structural hubs in the brain networks, namely, aggregated, distributed and connector hubs. Spatially, these distinct categories of hubs were primarily located in the default-mode system and additionally in the visual and limbic systems for aggregated hubs, in the frontoparietal system for distributed hubs, and in the sensorimotor and ventral attention systems for connector hubs. Importantly, these three categories of hubs exhibited various distinct characteristics, with the highest level of microstructural organization in the aggregated hubs, the largest wiring cost and topological vulnerability in the distributed hubs, and the highest functional associations and cognitive flexibility in the connector hubs, although they behaved better regarding these characteristics compared to non-hubs. Finally, all three categories of hub indices displayed high across-session spatial similarities and acted as a structural fingerprint with high predictive rates (100%, 100% and 84.2%) for individual identification. Collectively, our findings highlighted three categories of brain hubs with differential microstructural, functional and cognitive associations, which may shed light on the topological mechanisms of the human connectome.

# Introduction

The human brain works as a complex network to support various cognitive processes through information communication and integration between interconnected regions [1, 2, 3]. Using non-invasive diffusion magnetic resonance imaging tractography approaches, researchers have been able to reconstruct structural human brain networks at the macroscale [4, 5]. With a graph-theoretical network analysis framework, recent studies have suggested that structural brain networks contain sets of centrally embedded and topologically important hub nodes or regions, which are generally identified using various graph measures [6]. These brain hubs provide the anatomical underpinnings for the efficient transfer of information among regions [4, 5, 7] and consume high wiring cost and physiological energy [8, 9, 10]. Moreover, the abnormal topological properties of these hubs have been associated with a variety of neurological and psychiatric disorders, suggesting their vital roles to maintaining normal brain function [6, 11, 12, 13]. Together, such recent progress has highlighted the significances of structural hubs in understanding the biological mechanisms of the brain under healthy and diseased conditions.

Notably, structural hubs in human brain networks are usually identified as the nodes with high values of certain graph-based nodal metrics (e.g., degree centrality, closeness centrality, betweenness centrality or participant coefficient) [4, 5, 14] or combinations of several metrics [4, 15]. However, different nodal metrics capture different topological roles of nodes in brain networks. For example, for a given network node, the nodal degree centrality describes the number of connections linking to this node, the nodal betweenness centrality specifies the importance of information flow path, and the nodal participant coefficient describes the capability linking different network modules [16, 17]. Therefore, these structural hubs identified in previous studies are likely to represent distinct nodal roles in brain networks, but unfortunately, they are vaguely termed "hubs" in a general manner. From the empirical results, although different nodal metrics detect certain common hubs (e.g., the medial parietal cortex and the precuneus) [4, 5, 6, 15], there are some discrepancies in spatial locations even when using the same data set in one study: for instance, the inferior parietal cortex was identified as a hub using closeness centrality but not using betweenness centrality, and the superior temporal gyrus was identified as a hub by degree centrality but not by closeness centrality [4]. These facts from graph-based network theory and experimental observations raise an important question - whether there exist distinct category structural hubs with different topological roles in human brain networks, and if so, how they have different spatial distributions. Specifically, previous studies have demonstrated high-level microstructural organization [9, large wiring costs [8, 18] and functional associations [9, 19] in the structural hubs. Thus, whether distinct category hubs exhibit different characteristics remains to be elucidated. Answering these questions will greatly improve our understanding of the organizational principles and topological mechanisms of the human structural connectome.

To address these issues, we utilized diffusion MRI data (Dataset 1) to reconstruct individual structural brain networks and further estimated eight frequently used graph nodal metrics to characterize various aspects of the topological roles of each nodal region. Then, these nodal metrics were classified into different categories based on their spatial similarity, and structural brain hubs and hub indices were identified for each category. We further investigated the underlying microstructural organization, wiring cost, functional associations, cognitive flexibility and topological vulnerability of distinct category hubs. Moreover, based on a repeated scanning imaging dataset (Dataset 2), we compared the results of the classification of metrics and the spatial distribution of hub indices between two scanning sessions to evaluate their reliability and performed an individual identification analysis to assess the individuality of hub indices. Finally, validation analyses were conducted using different network constructions and analysis strategies as well as different diffusion imaging protocols (high angular resolution diffusion imaging, HARDI, Dataset 3).

## Results

**Three Categories of Structural Hubs in the Human Brain Networks**

*Similarity of Spatial Distribution of Network Nodal Metrics.* For each subject, we constructed individual structural brain networks and generated eight nodal centrality maps (Fig. 1). Visual examination indicated that several regions exhibited higher nodal centrality values (e.g., ranked in top 20%) in most of these centrality maps, which included the medial and lateral frontal and parietal regions and several subcortical regions, such as the putamen, the caudate and the thalamus (Fig. 2A). Further, Spearman's rank correlation analyses revealed a wide range of correlation values among these centrality maps (range: 0.38 ~ 0.93, Fig. 2B), indicating remarkably similar or different spatial distributions between specific network nodal metrics. Notably, all spatial correlations among these nodal centrality maps in the brain network were significantly ($Ps < 0.001$, Bonferroni corrected) lower than those in the randomized counterparts, except for the correlation between K-core centrality and subgraph centrality ($P = 0.0117$, uncorrected) (Fig. 2C), suggesting that different nodal metrics capture intrinsically distinct, unique organizational principles of brain networks.

*Three Categories of Structural Brain Hubs.* Using the agglomerative hierarchical clustering analysis on the group-averaged metric-by-metric correlation matrix, we classified the eight nodal metric maps into three categories (Fig. 3A and Fig. S1): i) subgraph centrality, K-core centrality, eigenvector centrality and closeness centrality; ii) page-rank centrality, betweenness centrality and degree centrality; and iii) participant coefficient. Within each category, the brain hubs were identified according to a hub index, and their topological and spatial positions were described (Fig. 3B and Fig. 4). Topologically, we observed that the three categories of brain hubs showed distinct features: closely aggregated hubs, widely distributed hubs and dispersed hubs that linked structural modules, which were thus defined as aggregated hubs (A-Hub), distributed hubs (D-Hub) and connector hubs (C-Hub), respectively (Fig. 3B). Using a functionally defined brain parcellation (Fig. S2), we showed convergent and divergent spatial distributions in functional systems among the three categories of structural hubs: commonly identified hub nodes in all three categories were primarily located at the default-mode system ($Ps < 0.005$, Bonferroni corrected), and hub nodes were additionally identified in the visual and limbic systems for aggregated hubs, in the frontoparietal systems for distributed hubs and in the sensorimotor and ventral attention systems for connector-hubs ($Ps < 0.01$, Bonferroni corrected) (Fig. 4A and 4B).

**Miscellaneous Characteristics of Three Categories of Structural Brain Hubs**

*Microstructural Organization and Wiring Cost of Structural Hubs.* Compared with the non-hubs, all three categories of brain hubs had significantly ($Ps < 0.001$, Bonferroni corrected) larger fractional anisotropy, mean diffusivity and axial diffusivity values, smaller radial diffusivity values, longer streamline lengths and higher streamline costs, except for the radial diffusivity of the connector hubs ($P = 0.0038$, uncorrected) (Fig. 5A). Among the three categories of hubs, the aggregated hubs exhibited the largest fractional anisotropy and axial diffusivity values and the longest streamline length, the distributed hubs exhibited the highest streamline cost, and the connector hubs exhibited the smallest fractional anisotropy, the smallest axial diffusivity, the largest radial diffusivity, the shortest streamline length and the lowest streamline cost ($Ps < 0.005$, Bonferroni corrected, Fig. 5B). These results together imply that all three categories of structural brain hubs tended to retain high-level microstructural organization and expensive wiring costs compared to the non-hubs but showed significant differences in these features among the categories of hubs.

*Functional Roles and Cognitive Flexibility of Structural Hubs.* Both the distributed hubs and the connector hubs had significantly ($Ps < 0.001$, Bonferroni corrected) higher functionally defined participant coefficients and more number of cognitive components than the non-hubs; in contrast, the aggregated hubs had no significantly different functional participation coefficient or number of cognitive components with the non-hubs (Fig. 5A). Furthermore, sorted in descending order for both functional participant coefficient or the number of cognitive components, the hubs followed connector hubs > distributed hubs > aggregated hubs ($Ps < 0.001$, Bonferroni corrected, Fig. 5B). These results indicate that both distributed and connector hubs play more crucial roles in functional integration among systems and contribute to larger cognitive flexibility than aggregated hubs.

*Topological Vulnerability of Structural Hubs.* To assess the effect of 'lesions' in the three categories of structural hubs on the overall topological vulnerability of brain networks, we performed simulation analyses in which network nodes were continuously removed in a manner of random failure or targeted attacks. As expected, the continuous attacks on all three categories of hubs had more dramatic effects on the brain network performance than on the random failure of nodes, as indicated by the significantly lower AUC of the largest component size and global efficiency ($Ps < 0.001$, Bonferroni corrected, Fig. 5A and Fig. 6). Of note, targeted attacks on distributed, aggregated and connector hubs resulted in reduced network performances in a descending order when the top 20% of nodes were removed ($Ps < 0.001$, Bonferroni corrected, Fig. 5B and Fig. 6). Together, our results suggest that all three categories of structural hubs are critical for maintaining global communication and the topological stability of the brain networks, while connector hubs are the most resilient to targeted attacks compared to the other categories.

**Highly Reliable Brain Hub Indices and Their Contributions to Individual Identification**

*High Classification Reliability of Metrics and Spatial Distribution of Hub Indices.* We classified eight metrics into three categories using Dataset 2 and found that the results of metric classification from two scanning sessions were entirely consistent (data not shown), suggesting that the classification of metrics and hub redefinition are highly reliable. Furthermore, the spatial distributions were significantly similar between scanning sessions for all three category hub indices (individual-level: $\rho = 0.77\pm0.03$, $0.73\pm0.02$ and $0.35\pm0.05$; group-level: $\rho = 0.99$, $0.99$ and $0.94$ for aggregated, distributed and connector hub indices, respectively) (Fig. S3), which suggests the high reliability of the spatial distribution for all three category hub indices.

*Hub Index Contributions to Individual Identification.* We implemented individual identification for each category of hub indices. From Session 1 to Session 2, the predictive rates of subject identity reached 100%, 100%, and 89.5% for aggregated hubs, distributed hubs, and connector hubs, respectively. The similar predictive rates from Session 2 to Session 1 were 100%, 100%, and 84.2%, respectively. These findings suggest that the spatial patterns of all three category hub indices are unique across individuals and could serve as a structural fingerprint for individual identification.

**Validation Results**

Our results were evaluated from three different aspects that involved different diffusion imaging protocols and fiber reconstructing algorithms, node definitions and hub selection thresholds. Our main findings were not affected by these factors, indicating the strong robustness of the three categories of brain hubs. For details, see Supplementary Results and Fig. S4-S9.

# Discussion

We defined three categories of structural brain hubs, namely aggregated, distributed, and connector hubs, with anatomically convergent and divergent spatial distributions in brain systems. Moreover, these distinct category brain hubs showed differential microstructural, functional and cognitive associations and topological vulnerability. Importantly, all three categories of structural hubs retained high reliability in both spatial locations and microstructural and functional characteristics across long-term repeated scans and can act as a structural fingerprint with high predictive rates for individual identification. To our knowledge, we demonstrated for the first time three categories of structural brain hubs with different topological roles and functional significances, which highlights organizational principles of human brain structural networks.

**The Classification of Graph-Nodal Metrics**

By performing hierarchical clustering analysis, the eight commonly used nodal metrics were classified into three categories according to their spatial distributions. The first category included the subgraph centrality, K-core centrality, eigenvector centrality and closeness centrality. Mathematically, these four metrics were designed to assess the position of a node in the topological space of the networks. For instance, the K-core measures the distance of a node that is far from the periphery of the network [20]. It is not surprising that the hubs identified by using these metrics were keen to topologically aggregate in the central of the networks, forming a structure at the higher level of the hierarchical architecture in the network topology. Thus, these hubs could correspond to the "sources" or "sinks" structure that work as an input or relay station for the whole networks [20, 21, 22, 23]. In contrast, the second category included page-rank centrality, betweenness centrality and degree centrality, which captures the capacities of global and local connection integration or position in the communication path of the networks. As expected, hubs of this category were evenly distributed in both central and periphery of the networks, and work as the 'router' for supporting the information transferring in both global and local levels [22, 24, 25, 26]. Moreover, the participant coefficient quantifies the level of a node to connect with different anatomical modules, and the PC hubs were naturally to be located at the adjacent borders of the anatomical lobes, which act as "connector" to facilitate the communication among anatomical modules underlying the functional parcellations [27]. Together, our classification for these commonly used nodal metrics was highly compatible with their definitions in graph theory, and the three categories of hubs further emphasized the specifically topological and anatomical characteristic of each metric.

**The Spatial Distribution of Three Categories of Structural Hubs**

The three categories of structural hubs shared common spatial distributions primarily in the default-mode system, including the posterior cingulate cortex/precuneus, the medial prefrontal cortices and the middle temporal cortices, which is in line with previous structural hub studies [4, 5, 6, 8, 11, 15]. The default-mode system has been proved to be a core system of the human brain network with a high cost of energy consumption [28, 29] and heavy regional cerebral blood flow [30] to support highly efficient information transfer [8, 18] and various cognitive process [31, 32]. Here, our results provided further evidence that the default-mode system is the central system of structural brain networks from different topological perspectives. Intriguingly, we found that different categories of hubs also exhibited unique distributions in several specific brain systems. For instance, aggregated hubs were more located in the visual and limbic systems. The visual system is a fundamental system for capturing information from the outside environment, which lies near the bottom of the visual processing hierarchy. It provides the original signal for further processing by other functional systems, which makes it function as a "source" structure in brain networks [33, 34, 35]. The limbic system supports a variety of functions, including emotion, motivation and behavior, and by its complex anatomical connections with both the association cortex and basal ganglia, it acts as a "relay station" [36, 37, 38]. Distributed hubs tended to be more evenly distributed in all functional systems. This even distribution characteristic might suggest their crucial roles in segregating and integrating information from separate parts of the whole-brain networks at both global and local levels [2, 6, 10]. Connector hubs were found more significantly located in the sensorimotor and ventral attention systems than the other categories of hubs. Although the sensorimotor and ventral attention systems have their own independent functions, their anatomical locations are precisely settled on the boundaries of brain lobes (e.g., the sensorimotor system at the central sulcus and the ventral attention system in the parieto-occipital sulcus and boundary of the cingulate/frontal cortices). Notably, the locations of the connector hubs reported in our study are consistent with a previous study [4]. Collectively, our results demonstrated the convergence and divergence of the anatomical distributions of the three categories of hubs, possibly suggesting a potential topological architecture of structural networks underlying brain functional systems.

**The Characteristics of the Three Categories of Structural Hubs**

As expected, all three categories of hubs showed higher level of microstructural organization than non-hubs, as indicated by the larger fractional anisotropy, mean diffusivity and axial diffusivity values, suggesting that these hubs are associated with regular fiber architecture, greater axonal diameter, larger packing densities and higher proportions of myelinated axons in white matter (WM) tracts [9, 39, 40]. These hubs were also keen to connect with distant fiber streamlines and to require greater wiring cost for building these topologically centralized hubs that facilitate communication with distant brain regions. [8, 9, 18]. Specifically, the aggregated hubs had the highest level of microstructural organization and the longest transmission distance, which might allow them to transfer information efficiently within whole-brain networks. The distributed hubs consumed the greatest wiring cost, which might be due to their dispersed involvement in the integration of both global and local communication. Nonetheless, these WM traits empower all three categories of hubs to maintain fast and long-distance communication with shorter transmission delays but larger physical consumption, which consequently facilitate synchronous information processing and increased signal transfer robustness during communication [9, 18, 41]. This phenomenon provides further experimental support for the existence of the cost-efficient trade-off of neural systems [10] and suggests that three categories hubs are all important components in this neural circuitry formation. Moreover, all three categories of hubs, especially the distributed hubs, had pronounced topological vulnerability as assessed by the "lesion" simulation, suggesting the core positions of all three categories of hubs in supporting the architectural organization and efficient information communication of brain structural networks [11]. Additionally, both distributed hubs and connector hubs exhibited significant higher functionally defined participant coefficients and larger cognitive flexibility than peripheral nodes, suggesting that these two categories of structural hubs are more involved in information integration among functional modules than underlie multiple cognitive functions than aggregate hubs [42, 43, 44, 45].

**The Robustness and Individually Uniqueness of the Three Categories of Structural Hubs**

We demonstrated that the spatial distributions were significantly similar between scanning sessions for all three categories of hub indices, indicating that the existence of three categories of hubs was not induced by "artifacts" and the three categories of hub indices are useful in reflecting the organizational characteristics of structural human brain networks. More importantly, our results also showed that the intra-subject individual variability of all three categories of hub indices are relatively smaller than the inter-subject variability, and they could serve as the connectome fingerprints for accurately identifying subjects from one another [46]. This suggests that the pattern of the three categories of hubs in the structural network is unique for each individual, which might be a crucial connectome basis for exploring the variance in individual behavior and implementing personalized medicine for neuropsychiatric illnesses [46, 47, 48].

**Limitations and Further Considerations**

There are several issues that warrant further considerations. First, we demonstrated that our main findings are reliable under different methodological choices, including imaging protocols, fiber reconstructing algorithms, brain parcellations and hub selective thresholds. Nevertheless, with the growth of neuroimaging techniques, newly developed methods, such as diffusion spectrum imaging [49] and multi-modal brain parcellations [50], may be taken into account in the future. Second, our analyses were performed based structural brain networks constructed from diffusion MRI data, which disabled exploration in the directions of the fiber tracks, resulting in the inability to map information flow in the structural brain networks. Future studies that are established using more advanced imaging techniques or data from postmortem brains [51] might provide opportunities to deepen our understanding of the directed topologies of the three categories of hubs. Third, we adopted WM diffusion indices to estimate the microstructural organization of the structural hubs in the current study; however, the accurate biophysical interpretation of these indices remains to be further clarified [52]. Future studies combining biophysical data from microscale cytoarchitectonics [53], myeloarchitectonics [54], chemoarchitectonics [55], and the metabolic level [9, 30] might better explain the association between the structural connectome and its material substrates. Finally, previous studies have suggested a strong nexus between brain hubs and neuropsychiatric disorders [11], and our "lesion" simulation also demonstrated a relatively different effect on attacking different categories of structural hubs. Future studies using data corresponding to disease states are desirable to ascertain the specific associations among distinct categories of hubs and brain disorders, which would extend our insight into the pathologies of neuropsychiatric disorders, and in turn, allow for a better understanding of the biological meaning of the diverse topology of different brain hubs.

## Materials and Methods

### Data Overview and Participants

Three imaging data sets were included in this study (Table 1): a principal dataset of 146 participants with structural MRI, diffusion tensor imaging (DTI) and resting-state functional MRI (R-fMRI) data (Dataset 1), a test-retest dataset of 57 participants with structural MRI and DTI data (Dataset 2) and a validation dataset of 38 participants with structural MRI and HARDI data (Dataset 3). All these participants were right-handed and had no history of neurological or psychiatric disorders. Written informed consent was obtained from each participant. The study designs from Dataset 1 and Dataset 2 were approved by the Institutional Review Board of the State Key Laboratory of Cognitive Neuroscience and Learning at Beijing Normal University, and the study design for Dataset 3 was approved by the Institutional Review Board of the WU-Minn Human Connectome Project (HCP) [56]. Table 1 shows the demographics of all participants and imaging modalities used in this study. Notably, Dataset 1 was used for the principal network analyses in this study, involving constructing structural brain networks, identifying structural hubs and exploring miscellaneous characteristics of structural hubs (e.g., microstructural organization, wiring cost, functional associations and topological vulnerability); Dataset 2 was used for the reliability analysis for structural network hubs and individual identification; and Dataset 3 was used for the validation of the main results obtained from Dataset 1. The scanning parameters of these three data sets were described in the Supplementary Methods.

### Constructing Individual Structural Brain Networks

Nodes and edges are the two basic elements of a network. Here, the structural and diffusion imaging data of Dataset 1 were used to construct structural human brain networks. The relevant procedures of network construction were introduced in our previous works [5, 57, 58] (Fig. 1) and are briefly described as follows.

*i) Definition of Network Nodes.* The procedure for defining network nodes was implemented by using the SPM8 package (https://www.fil.ion.ucl.ac.uk/spm/software/spm8/). Briefly, for each individual, the T1-weighted image was first co-registered to the averaged b0 image in the native diffusion space using a linear transformation. The co-registered T1-weighted image was then segmented into gray matter, white matter and cerebrospinal fluid by using a unified segmentation algorithm. The resultant images were further nonlinearly registered into the Montreal Neurological Institute space, and the transformation matrix was estimated. Finally, the inversed transformation matrix was used to warp a predefined brain parcellation with 1,024 regions of interest from the standard space to the native diffusion space. Discrete labeling values in the parcellation were preserved by the use of a nearest-neighbor interpolation method. Notably, the brain parcellation was generated by randomly subdividing the automated anatomical labeling atlas into 1,024 cortical and subcortical regions of equal size [59], which allowed capturing both major tracts and forking U-fibers. Thus, for each individual, we obtained 1,024 brain regions, each representing a network node.

*ii) Definition of Network Edges.* The procedures for defining network edges were mainly based on the whole-brain fiber bundles, which were reconstructed using the deterministic tractography method. Briefly, each individual diffusion weighted image was first preprocessed (eddy current and motion artifact correction) and aligned to an averaged b0 image by an affine transformation using the FMRIB's Diffusion Toolbox of FSL (Version 5.0; https://www.fmrib.ox.ac.uk/fsl). Then, the Fiber Assignment by Continuous Tracking (FACT) algorithm [60] was performed to reconstruct all WM bundles in the brain using the DTIstudio package (version 3.0.3). Here, fiber tracking was computed by seeding each voxel with a fractional anisotropy value greater than 0.2. This fiber-tracking procedure was terminated at voxels where fractional anisotropy < 0.2 or if the turning angle between adjacent steps was greater than 45°. Using this procedure, tens of thousands of streamlines were generated to etch out all of the major WM tracts. Two nodes were considered to be structurally connected if there were streamlines with end points located in these two regions. To this end, for each subject, we obtained a binary WM network, and subsequent analyses were conducted at an individual level, unless specifically mentioned.

**Identifying Categories of Structural Brain Hubs**

To identify structural hubs from the brain networks and to further ascertain whether they could be classified into different categories, we utilized eight widely used graph-based metrics to quantify nodal roles in the brain networks and performed further hierarchical clustering analysis. All these analyses were performed using the GRETNA toolkit (https://www.nitrc.org/projects/gretna/) [61], the MatlabBGL package (https://www.cs.purdue.edu/homes/dgleich/packages/matlab_bgl/), the brain connectivity toolbox (BCT, https://sites.google.com/site/bctnet) and our in-house Matlab codes. The procedures in detail are described as follows.

*i) Graph Nodal Metrics.* We studied eight graph-based centrality metrics, including betweenness centrality (Bc), closeness centrality (Cc), degree centrality (Dc), eigenvector centrality (Ec), K-core centrality (Kc), participant coefficient (Pc), page-rank centrality (Pr), and subgraph centrality (Sc) (Fig. 1, Table S1). These metrics capture different topological roles of network nodes and have been widely adopted in previous brain network studies [4, 5, 14, 15, 43, 62, 63, 64]. Notably, these nodal metrics result in different ranges of values while they are computed in a network. For comparing across metrics, we transferred the original values of each metric into corresponding temporary ranking scores from 1 to 1,024. Regarding the nodes with the same value, we computed the mean value of their temporary ranking scores, resulting in the final nodal ranking scores. For a given metric, the nodes with higher ranking scores correspond to nodes with higher topological roles in a network.

*ii) Spatial Similarity among Metrics.* To investigate the similarity of spatial distributions among these eight nodal metrics, for each subject we first computed the Spearman's rank correlation ($\rho$), between every pair of metrics across nodes, resulting in an eight-by-eight correlation matrix. To further determine whether the spatial similarity between any pairs of metrics in brain networks occur by chance, we compared these correlation matrices of all the individuals with those derived from 100 random networks, which were generated by using Maslov's wiring algorithm, retaining the same number of nodes, number of edges and degree distribution as real brain networks [65]. Then, the Z-scores were estimated to quantify the differences between the $\rho$ values of brain networks and random counterparts:

$$z_{a,b} = (\rho_{a,b} - \mu_{a,b})/\sigma_{a,b}, \quad (1)$$

where $\rho_{a,b}$ is the Spearman's rank correlation coefficient between metric $a$ and metric $b$ in the brain network, and $\mu_{a,b}$ and $\sigma_{a,b}$ are the mean and the standard deviation of Spearman's rank correlation coefficients between the two metrics in the random networks, respectively. Finally, for every pair of metrics we performed one-sample t-tests across individuals to determine whether these Z-score values were significantly different from zero, and Bonferroni correction was used for multiple comparisons (i.e., $P < 0.05/28$).

*iii) Hierarchical Clustering Analysis and Categories of Network Hubs.* To determine whether the eight nodal metrics can be classified into different categories, we performed the following hierarchical clustering analysis. Briefly, for each individual the metric-by-metric Spearman's correlation matrix was first transformed to Fisher's z matrix using Fisher's $\rho$-to-z transformation to improve normality. The Fisher's *z* matrices were averaged across individuals and further inverse-transformed to generate a new group-based correlation matrix. Then, we obtained a dissimilarity matrix by subtracting the correlation values from 1 and generated an agglomerative hierarchical clustering tree based on the single linkage algorithm using weighted average distance metric. Thus, eight nodal metrics were classified to different categories. To determine a proper category number, we employed a stability analysis procedure [35, 66] in which a hierarchical clustering analysis was performed on the Spearman's correlation matrices that were obtained by randomly choosing 5% of all subjects 1000 times. Based on this procedure, the eight nodal metrics were classified into three categories (for details, see Results) in which the category number was significantly stable and simultaneously ensured larger category-assigning differences of nodal metrics compared to null models (for details, see Supplementary Methods, Fig. S1). Finally, for each category, we identified the brain network hubs using a hub index, which was defined as the mean ranking score of metrics in this category. The nodes with the top 20% of hub index were identified as hubs, and the remaining considered as non-hubs (Note: two additional thresholds, 15% and 25%, were used for the validation analyses). To display the distribution of each category of hubs, all hub and non-hub nodes were unfolded in the topological space using "spring model" layouts based on the "fdp" algorithm (https://www.graphviz.org).

*iv) Distributions of Structural Hubs in Functional Brain Systems.* To examine whether and how different category structural hubs are associated with the brain's functional systems, we performed a functional network modularity analysis and further calculated the proportions of each category of hubs distributed in each system. Briefly, we first built a group-based functional brain network at a voxel-level using the R-fMRI data of 146 participants from Dataset 1 and then identified functionally connected modules using a graph-based network modularity analysis (for details, see Supplementary Methods). Seven major functional subdivisions were identified, including the default-mode, visual, frontoparietal, sensorimotor, limbic, dorsal attention and ventral attention systems (Fig. S2), and this subdivision was largely compatible with previous functional brain network studies [35, 42, 67]. For each category of hub, we computed their proportions that belonged to different functional systems. Finally, we performed nonparametric permutation-based paired tests (20,000 permutations, the same below) across individuals with Bonferroni correction to evaluate the significance levels of the differences in the proportions of hubs among the functional systems or among the categories of hubs (i.e., $P < 0.05/84$).

**Characterizing Microstructural, Functional and Cognitive Associations of Structural Hubs**

To determine whether different category structural hubs exhibited common and distinct properties, we systematically explored their microstructural organization, wiring cost, functional roles, cognitive flexibility and topological vulnerability.

*i) Microstructural Organization and Wiring Cost of Structural Hubs.* We explored the underlying microstructural organization and wiring cost of different category hubs using the four WM diffusion indices and two WM wiring cost indices, respectively. For a given network edge, we first computed the four diffusion indices, including fractional anisotropy, which reflects the degree of anisotropy of a diffusion process; axial diffusivity, which estimates the level of diffusion in the direction of the first eigenvector used to describe the level of local fiber orientation; radial diffusivity, which reflects the amount of diffusion perpendicular to the first eigenvector and specifies the level of myelination of the WM; and mean diffusivity, which assesses the total level of diffusion [39, 68]. These diffusion indices were estimated by averaging the values across the WM voxels that the streamlines passed through, and they reflect the different aspects of the diffusion properties of WM tissues. Recent studies have suggested that the diffusion indices are approximately associated with the microstructural organization of WM tracts, such as axonal membrane or myelin [40]. Then, we computed the two WM wiring cost indices, including the following: streamline length, which captures the average length of all reconstructed streamlines in the network edge; and streamline cost, which represents the total streamline length in the network edge [8, 10, 41]. For a given network node, we obtained its diffusion and cost indices by computing the mean value of the edges that this node links. Next, the diffusion and wiring cost indices were averaged across hub and non-hub nodes, respectively. Notably, for isolated nodes, we cannot estimate their WM indices; therefore, when calculating the averaged diffusion and wiring cost indices across hub and non-hub nodes, the isolated nodes were ignored. Finally, we compared the diffusion and wiring cost indices between hubs and non-hubs (i.e., $P < 0.05/18$) or among different category hubs (i.e., $P < 0.05/18$) across individuals by using the nonparametric permutation-based paired tests with Bonferroni correction.

*ii) Functional Roles and Cognitive Flexibility of Structural Hubs.* We further studied whether different category structural hubs play distinct roles in the functional integrity of brain networks and whether they contribute to different cognitive flexibility underlying multiple cognitive functions. To do so, we first computed the functional participant coefficient at each voxel according to the modular architecture derived from the above-mentioned group-based functional brain network (for details, see Supplementary

Methods). For a given type of node (i.e., hub or non-hub nodes) in the structural brain network, we computed its functional participant coefficients by averaging the participant coefficients across all voxels belonging to the corresponding nodal type. Second, we obtained the brain map of the functional flexibility from Yeo et al. [44] (https://surfer.nmr.mgh.harvard.edu/fswiki/BrainmapOntology_Yeo2015), in which a cognitive component number was assigned to each voxel. Likewise, for a given nodal type in the structural network, we calculated its functional flexibility by averaging the cognitive component numbers of all voxels belonging to the corresponding nodal type. Finally, we performed nonparametric permutation-based paired tests with Bonferroni correction to evaluate the significance levels of the differences in either the functional participant coefficient or the cognitive component number between hubs and non-hubs (i.e., $P < 0.05/6$) or among categories of hubs (i.e., $P < 0.05/6$).

*iii) Topological Vulnerability of Structural Hubs.* We estimated the topological vulnerability of different categories of structural hubs using the following nodal "lesion" simulation procedure [14, 69]. Briefly, for each category of hubs, we first performed targeted attacks on individual structural networks by removing the nodes one-by-one according to the descending order of hub indices and then measured the changes in the global efficiency and the size of the largest connected component of the networks. We also performed a random failure procedure in which brain nodes were continuously and randomly removed from individual networks 100 times and recomputed the averaged two measures of the resultant networks. Notably, to ensure that the curves from different individual networks were comparable, for each curve, we divided all the values of this curve by the value of its first point to yield the normalized curve. Then, for each individual network, we calculated the area under the top 20% curves (AUC) of both the largest component size and the global network efficiency under targeted attacks and random failures. A smaller AUC represents a faster decrease in global network performance in response to nodal removal. Finally, we evaluated the differences in the AUC of the largest component size or the global network efficiency between when under targeted attacks and when under random failure (i.e., $P < 0.05/6$) or among the three categories of hubs under targeted attacks (i.e., $P < 0.05/6$) using nonparametric permutation-based paired tests with Bonferroni correction.

**Reliability of Structural Brain Hubs and Individual Identification Analyses**

To determine whether the classification of nodal metrics and the spatial distribution of structural brain hub indices are reliable and whether each category hub index can contribute to individual identification during repeated scans, we performed the following analyses using the imaging data of Dataset 2.

*i) Reliability Analysis.* For each subject, we used the DTI data from Dataset 2 to construct two structural brain networks corresponding to two scanning sessions to classify their eight metrics into three categories and to then compute their hub indices for each category. The network construction and analysis procedures were identical to those used for Dataset 1. To evaluate the reliability of the classification of metrics and the redefinition of hubs, we determined whether the results of the classification between two scanning sessions were consistent. We further calculated the Spearman's correlation coefficients for each category of hub indices across nodes between paired individual networks in two scanning sessions and also estimated group-averaged correlations between two sessions, to assess the reliability of the spatial distribution of hub indices at both the individual and group level.

*ii) Individual Identification Using Structural Hub Indices.* To explore whether the spatial patterns of different categories of hub indices contribute to individuality, we performed the following individual identification analysis. This procedure was originally proposed to identify individuals based on the brain's functional connectivity matrix [46], but it was modified here by incorporating nodal hub indices of brain networks. Briefly, for each category of hub indices, we first selected an individual from Session 1 as a reference and then calculated the Spearman's correlation coefficient across nodes between this reference and every subject in Session 2. Then, we determined whether the reference itself retained the maximum correlation value among all individuals in Session 2; if so, we defined that the prediction succeeded, and otherwise that it did not. Using this procedure, we repeated the above analysis for each individual from Session 1 and calculated the predictive rate for each category of hub indices. We also performed the predictive analysis from Session 2 to Session 1.

**Validation Analysis**

To determine whether our findings are robust for use under different diffusion imaging protocols, reconstructing algorithms of fiber pathways, node definitions, and hub selection thresholds, we implemented comprehensive validation analyses. For details, see Supplementary Methods.

## Acknowledgments

We would like to thank Drs Xuhong Liao, Miao Cao, Zhengjia Dai and Yuhan Chen for their insightful suggestions and Dr. Ruiwang Huang for kind help in the collection of MRI data. This work was supported by the Natural Science Foundation of China (Grant Nos. 81401479, 91432115, 81225012 and 31221003), the Beijing Natural Science Foundation (Grant No. Z151100003915082), and the Fundamental Research Funds for the Central Universities (Grant No. 2015KJJCA13). Data were provided [in part] by the Human Connectome Project, WU-Minn Consortium (Principal Investigators: David Van Essen and Kamil Ugurbil; 1U54MH091657) funded by the 16 NIH Institutes and Centers that support the NIH Blueprint for Neuroscience Research; and by the McDonnell Center for Systems Neuroscience at Washington University.

## Disclosures

The authors have no conflict of interest to declare.

**Table 1.** Datasets and Demographics

|  | **Dataset 1 (n=146)** | **Dataset 2 (n=57)** | **Dataset 3 (n=38)** |
| --- | --- | --- | --- |
| Gender (Male/Female) | 70/76 | 30/27 | 17/21 |
| Age (years) | 19-30 (22.68±2.24) | 19-30 (23.05±2.29) | 22-35[a] |
| MRI Modality | T1, DTI, R-fMRI | T1, DTI | T1, HARDI |

[a] For each subject, accurate year of age was not provided in the HCP dataset.

**Figure Legends**

**Figure 1. A Flowchart of the Network Construction and Graph Nodal Metric Estimation for Each Participant.** (1) After rigidly co-registering to the averaged b0 image, the native space T1 image was nonlinearly transformed to the ICBM 152 T1 template in the MNI space, resulting in a transformation matrix. (2) The inversed transformation matrix was utilized to warp the parcellation from the MNI space to the native space. (3) In terms of the parcellation and the results of deterministic tractography in the native space, the WM network was constructed. (4) Eight nodal metrics were estimated based on the individual WM networks. (5) The metrics were finally converted to normalized ranking scores.

**Figure 2. The Spatial Distributions of Eight Graph-Nodal Metrics and Their Spatial Similarities.** (A) The group-level centrality map for each graph-nodal metric was obtained by averaging the rank maps across individuals. The color of the surface represents the top percent for a given node in descending order of rank values. Notably, after obtaining the top percent maps, all results were smoothed for better visualization with full width half maximum (FWHM) = 2 mm. All eight smoothed group-level top percent of metrics were mapped to the ICBM152 brain surface template in MNI space. (B) The Spearman's rank correlations across nodes were estimated to represent the spatial similarities among nodal metrics. The order of metrics was arranged according to the following hierarchical clustering analysis to display the spatial similarities and dissimilarities among metrics. (C) The spatial similarities among nodal metrics were compared to a null model. The lower triangular matrix shows the $t$ values that represent the difference in spatial similarities among nodal metrics between those in the brain network and 100 random networks. The upper triangular matrix represents the significant level of $t$ values.

**Figure 3. Three Categories of Hubs.** (A) The group-averaged map of Spearman's correlations among the eight nodal metrics and the agglomerative hierarchical clustering tree generated from the map. The red, blue and green solid lines show the results of the classification, indicating the three categories of metrics, corresponding to the three categories of metrics used to identify the following aggregated hubs, distributed hubs and connector hubs. (B) The distributions of hubs from a representative subject in the topological space. Notably, the network layouts were generated using the "fdp" algorithm in NetworkX.

**Figure 4. The Distributions of Three Categories Hubs in Seven Functional Systems.** (A) The spatial distributions of group-level hub indices mapped on a brain surface (FWHM = 2 mm). (B) The proportions of the three categories of hubs in seven functional systems defined by our combined R-fMRI data. The red, blue and green solid lines indicate aggregated, distributed and connector hubs, respectively. Each color on the brain surface specifies a corresponding functional system.

**Figure 5. The Miscellaneous Characteristics of the Three Categories of Hubs.** (A) Comparisons of miscellaneous characteristics (e.g., microstructural organization, wiring cost, functional association, cognitive flexibility and topological vulnerability) between hubs and non-hubs for each category of hubs. Bonferroni corrections were performed for each block. (B) Comparisons of these characteristics among the three categories of hubs. The radar map shows the differences in the mean characteristic indices among the three categories of hubs; for each characteristic, the mean indices of the three categories of hubs were normalized from 0 to 1, where the minimal mean index was assigned as 0, and the maximal mean index was specified as 1. For each box plot, the bottoms and tops of the boxes indicate the first and third quartiles of the corresponding indices across individuals, the band inside the box represents the median, and the whiskers specify the 1.5 interquartile range (IQR) of the lower and upper quartiles.

**Figure 6. Targeted Attacks on the Three Categories of Hubs and Random Failure.** The descending curves of the largest component size (left plot) and global efficiency (right plot) for targeted attacks of the top 25% hub indices and random failure. The red, blue and green solid lines represent the mean curves for the targeted attacks on the aggregated, distributed and connect hubs across individuals, respectively; and the gray line represents the mean curve of random failure. The dashed areas represent the corresponding ±95% confidence intervals.

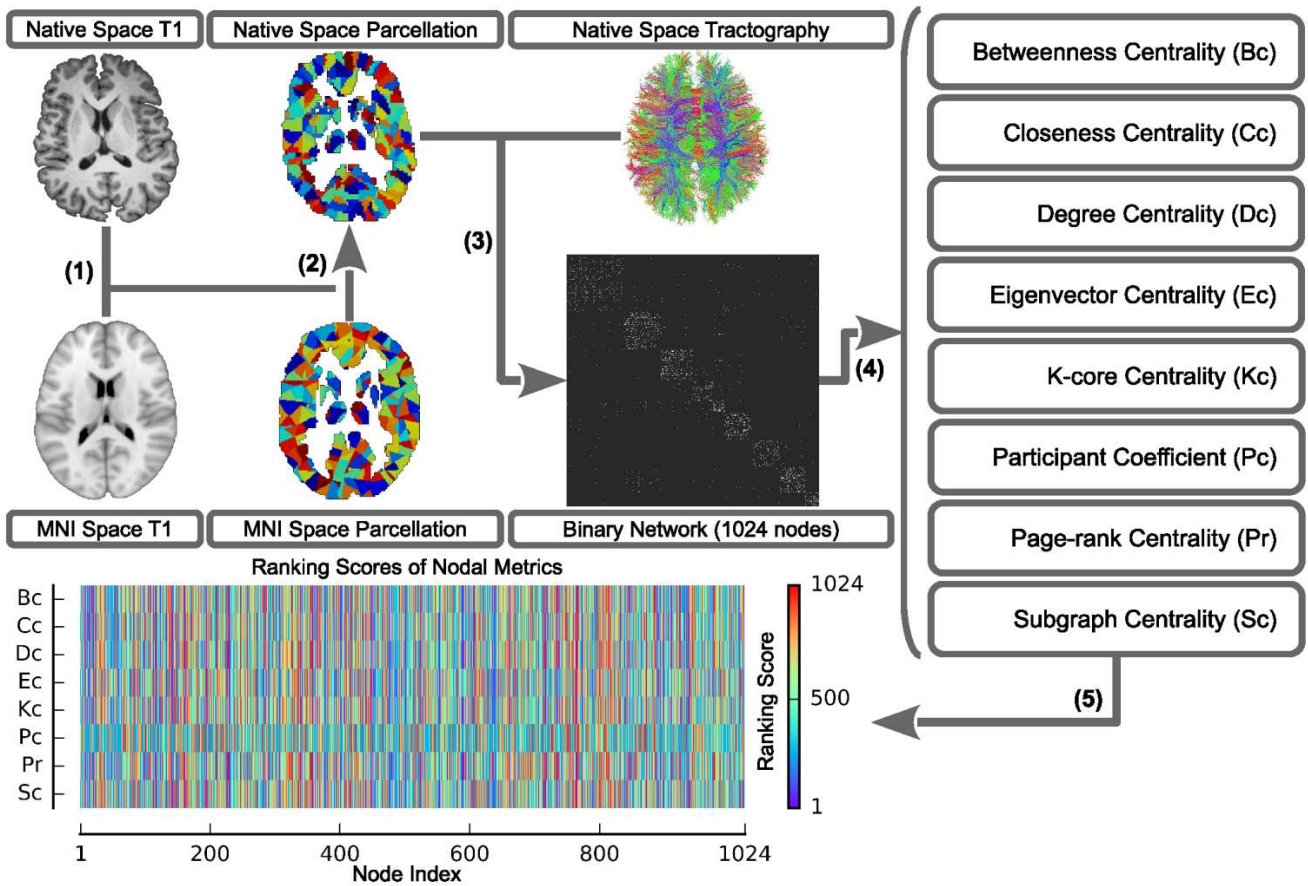

**Figure 1. A Flowchart of the Network Construction and Graph Nodal Metric Estimation for Each Participant.**

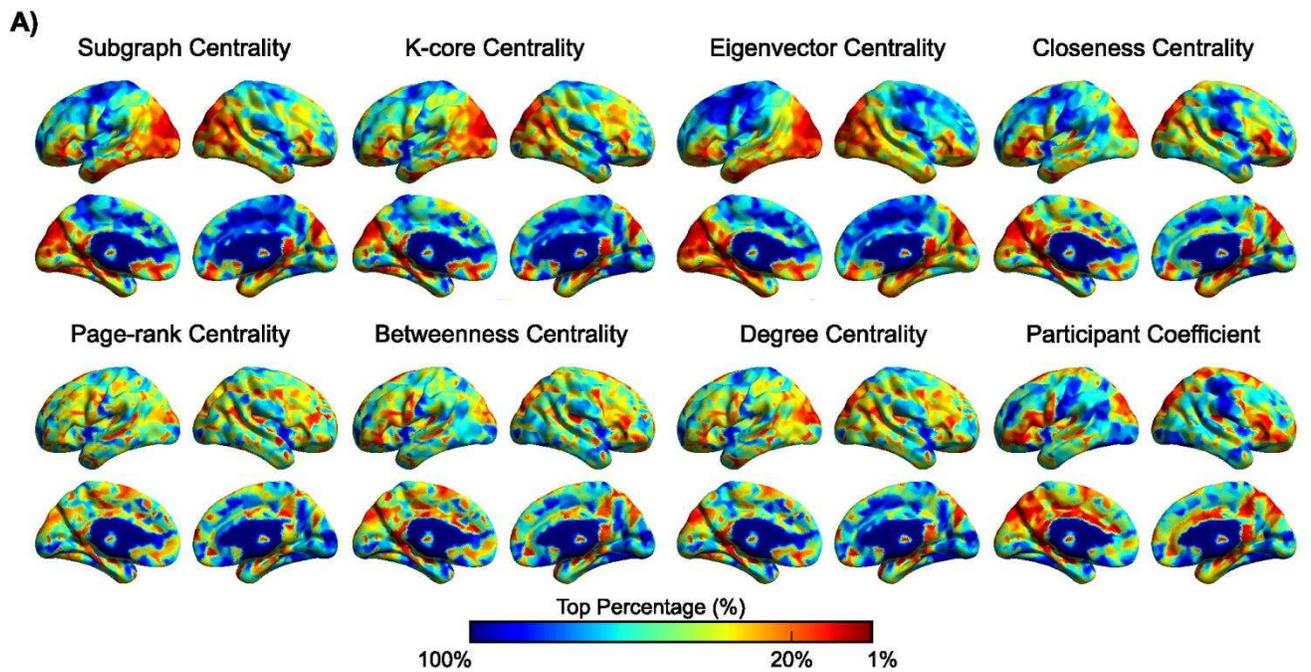
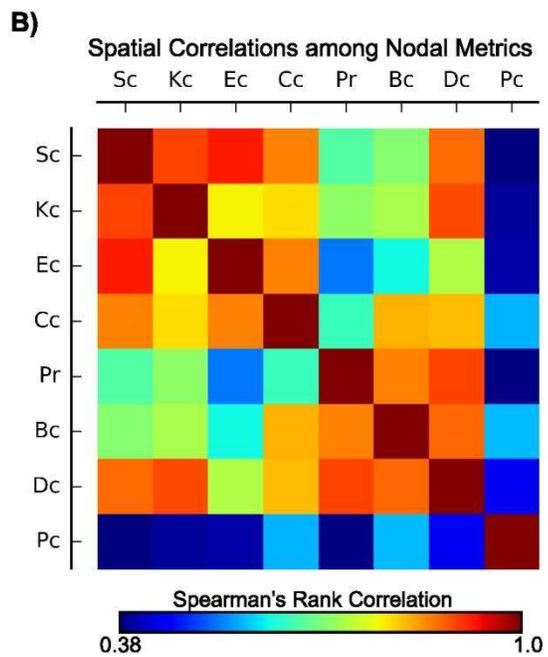
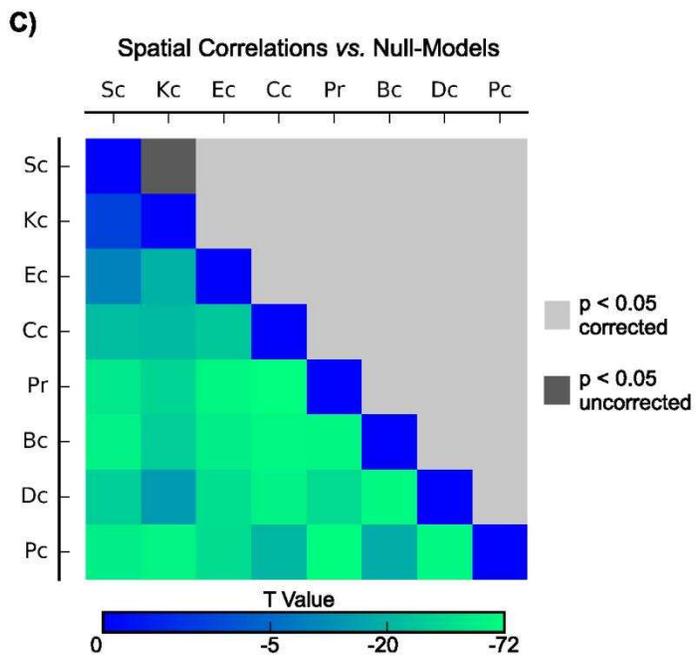

**Figure 2. The Spatial Distributions of Eight Graph-Nodal Metrics and Their Spatial Similarities.**

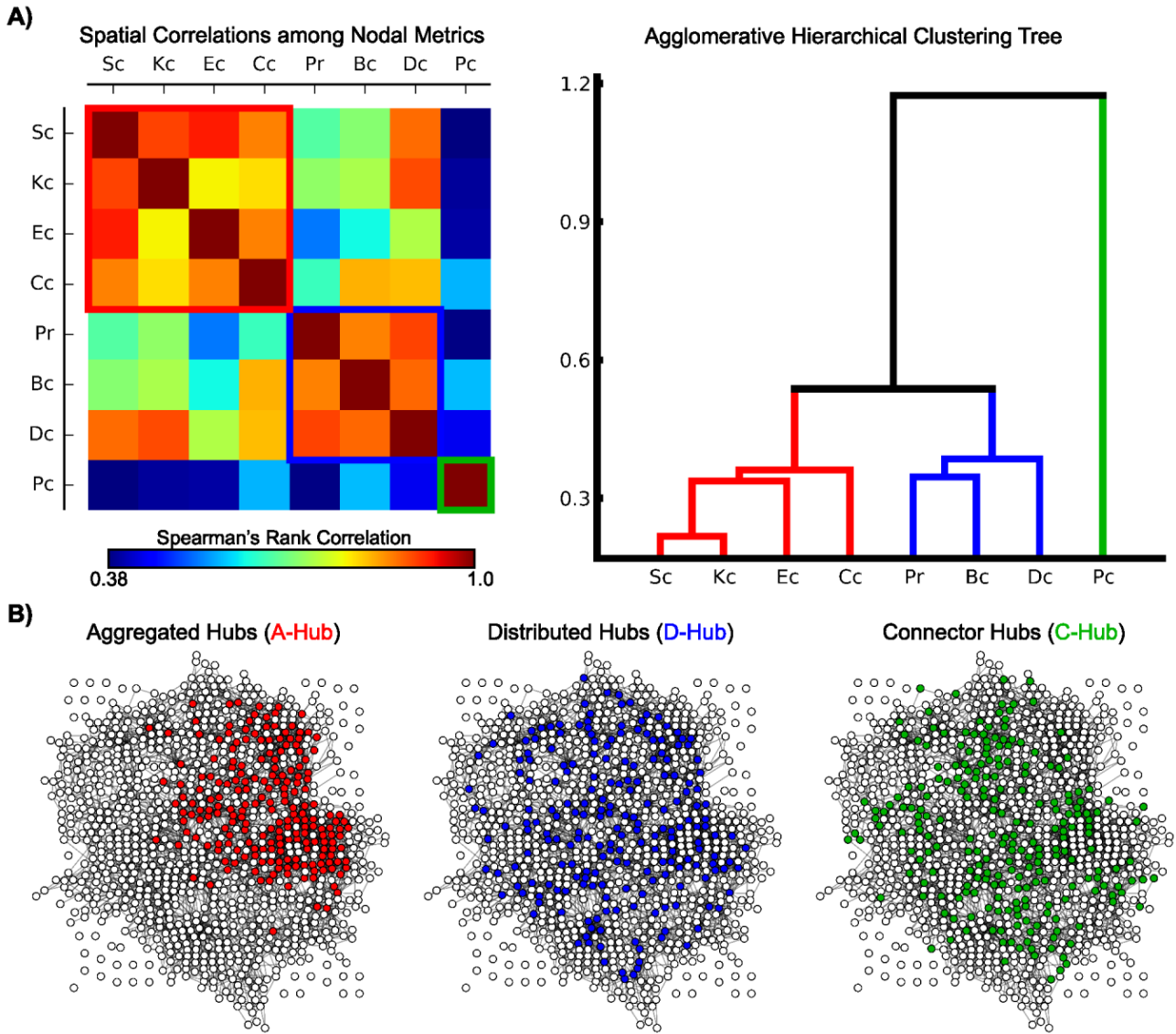

**Figure 3. Three Categories of Hubs.**

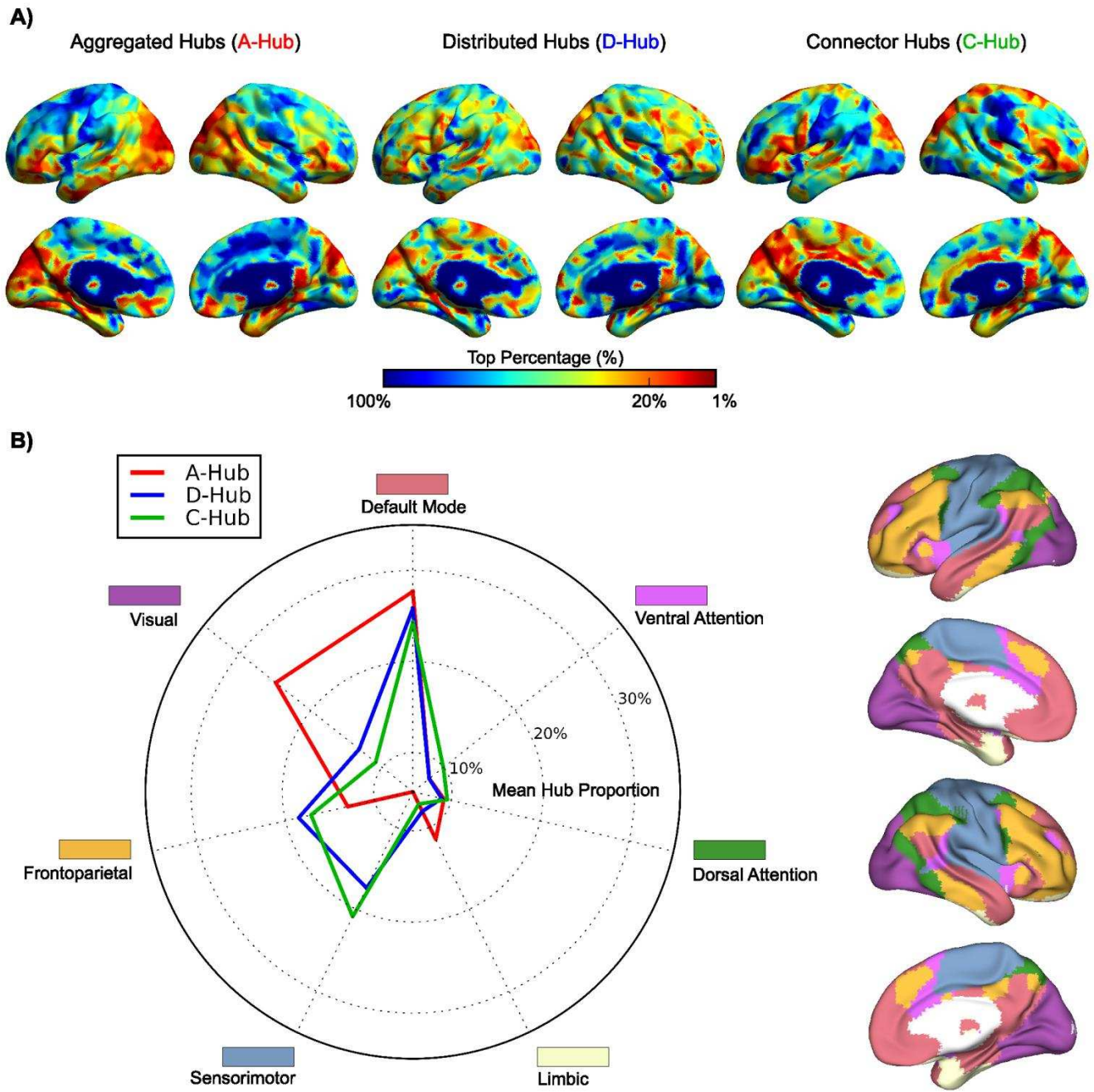

**Figure 4. The Distributions of Three Categories Hubs in Seven Functional Systems.**

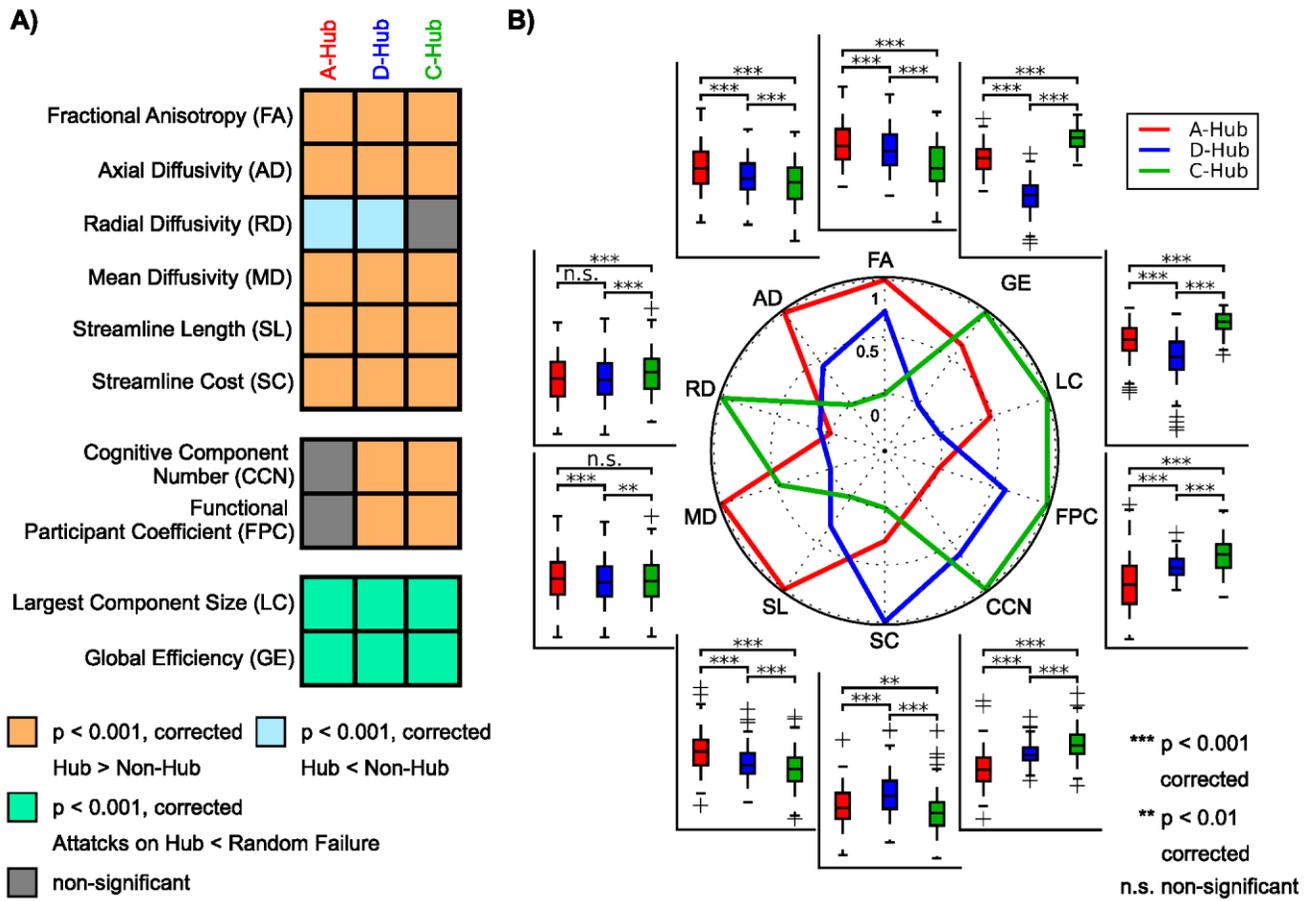

**Figure 5. The Miscellaneous Characteristics of the Three Categories of Hubs.**

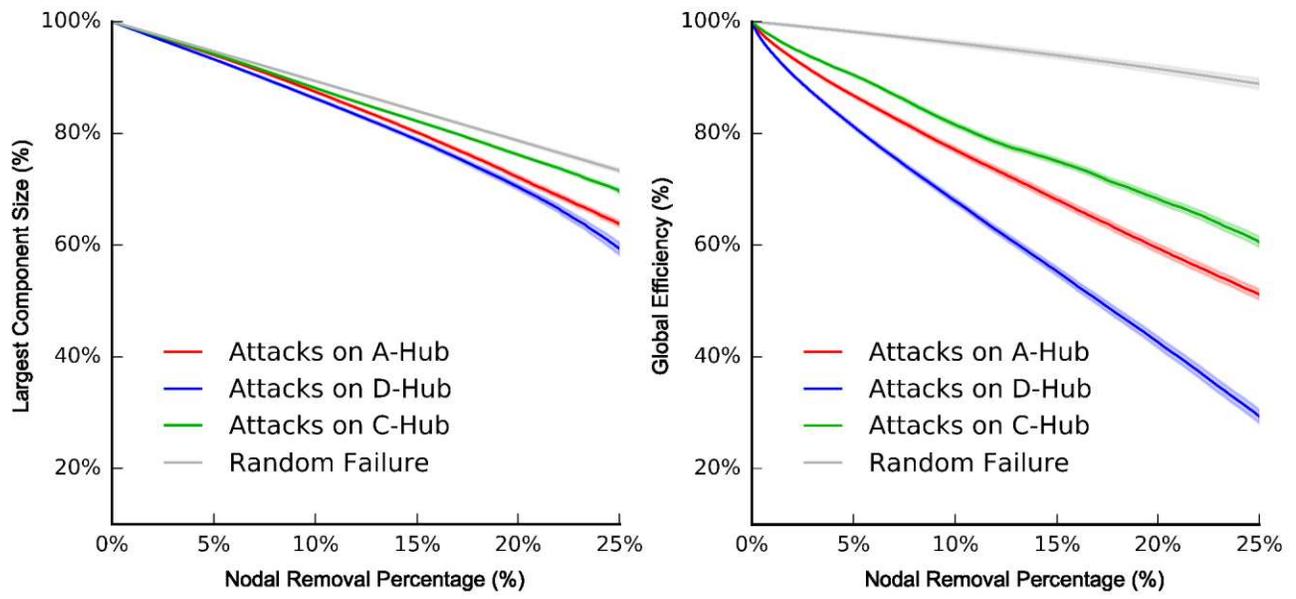

**Figure 6. Targeted Attacks on the Three Categories of Hubs and Random Failure.**

# Supplementary Information

## Supplemental Methods

**MRI Acquisition Procedures**

*Dataset 1:* The principal data set was selected from the Connectivity-based Brain Imaging Research Database (C-BIRD) at Beijing Normal University. All MRI data (structural MRI, DTI and R-fMRI) were acquired using a 3.0 T Siemens Trio Tim scanner (Siemens Medical Systems, Erlangen, Germany) with a 12-channel phased-array head coil in the Imaging Center for Brain Research, Beijing Normal University. The MR imaging procedures were as follows: *i) Structural MRI.* T1-weighted, sagittal 3D magnetization prepared rapid gradient echo (MP-RAGE) sequence, repetition time (TR) = 2,530 ms, echo time (TE) = 3.39 ms, inversion time (TI) = 1,100 ms, flip angle = 7°, matrix = 256 × 256, field of view (FOV) = 256 mm × 256 mm, slice thickness = 1.33 mm, voxel size = 1 mm × 1 mm × 1.33 mm and 144 sagittal slices covering the whole brain. *ii) DTI.* Single-shot twice-refocused spin-echo diffusion echo-planar imaging sequence, TR = 8,000 ms, TE = 89 ms, 30 non-linear diffusion directions with b = 1,000 s/mm$^2$ and an additional volume with b = 0 s/mm$^2$, number of excitation = 2, matrix = 128 × 128, FOV = 282 mm × 282 mm, 2.2 mm slice thickness, voxel size = 2.2 mm × 2.2 mm × 2.2 mm, bandwidth = 1,562 Hz/pixel, and 62 transverse slices without gap covering the whole brain. *iii) R-fMRI.* Echo-planar imaging sequence (EPI), TR = 2,000 ms, TE = 30 ms, flip angle = 90°, matrix = 64 × 64, FOV = 200 mm × 200 mm, 3.5 mm slice thickness, voxel size = 3.1 mm × 3.1 mm × 3.5 mm, 33 transverse slices with 0.7 mm gap covering the whole brain, and volume number = 200. This scan lasted for 6 min and 40 seconds. During the scan, the participants were instructed to rest and relax with their eyes closed and to refrain from falling asleep.

*Dataset 2:* The test-retest data set, including structural MRI and DTI data, was also from the C-BIRD. Notably, the participants in the Dataset 2 were scanned twice at an interval of approximate 6 weeks (40.94 ± 4.51 days) and had participated in the Consortium for Reliability and Reproducibility (CoRR) dataset (http://fcon_1000.projects.nitrc.org/indi/CoRR/html/bnu_1.html) [1]. The scanning parameters were identical to those of Dataset 1.

*Dataset 3:* The validation data set, including structural MRI and HARDI data, was selected from the WU-Minn HCP (https://db.humanconnectome.org, "unrelated 40 subjects") [2]. The original data set included imaging data for 40 healthy participants (Q1 and Q2 release), but two participants (subject ID: 209733 and 528446) were excluded because of structural brain abnormalities (https://www.humanconnectome.org/documentation/S500). The MRI data were acquired on an HCP's custom 3.0 T Siemens Skyra scanner using a 32-channel head coil at Washington University. The MR imaging procedures were as follows: *i) Structural MRI.* T1-weighted, sagittal 3D MP-RAGE sequence, TR = 2,400 ms, TE = 2.14 ms, TI = 1,000 ms, flip angle = 8°, matrix = 320 × 320, FOV = 224 mm × 224 mm, slice thickness = 0.7 mm, voxel size = 0.7 mm × 0.7 mm × 0.7 mm and 256 sagittal slices in a single slab [3]. *ii) HARDI.* A single-shot 2D spin-echo multiband EPI sequence, TR = 5,520 ms, TE = 89.5 ms, 270 diffusion directions with diffusion weighting 1,000, 2,000, or 3,000 s/mm² and 18 additional volumed with b = 0 s/mm², matrix = 144 × 168, FOV = 210 mm × 180 mm, 1.25 mm slice thickness, voxel size = 1.25 mm × 1.25 mm × 1.25 mm, 111 transverse slices without gap covering the whole brain [4].

**Identifying Categories of Structural Brain Hubs**

*Hierarchical Clustering Analysis and Categories of Network Hubs.* To determine the suitable category number for the hierarchical clustering analysis (HCA), we performed the following analyses: a stability analysis to test the stability of the classification results across populations and a null-model analysis to assess the organizational uniqueness of the classification results for each classified category. In the first analysis, the group-averaged metric-to-metric spatial similarity matrix was first computed by averaging the similarity matrices across all individuals and the HCA classification results (i.e., 2 to 7 categories) of this matrix were referred to as the real assignment. Then, we randomly selected 5% of individuals to generate a new averaged similarity matrix and performed HCA to generate the classification results for this matrix using different category numbers (the bootstrapping assignment). For each category number, an instability value was estimated as the disagreement of the classification results between the real assignment and the bootstrapping assignment by using the following formula [5, 6]:

$$E_c(\boldsymbol{BA}, \boldsymbol{RA}) = \min_{l \in BA} \frac{1}{N} \sum_{i=1}^{N} \delta(l_i \neq RA_i)$$

where $E_c$ specifies the instability value of the category number $c$, $\boldsymbol{BA}$ represents all possible assigning labels for the bootstrapping assignment, $\boldsymbol{RA}$ is one of the possible assigning labels for the real assignment, $l$ specifies one of the possible labels for the bootstrapping assignment, $N$ is the number of graph-nodal metrics, $l_i$ and $RA_i$ are the assigning labels of metric $i$ for the bootstrapping assignment and the real assignment, respectively, and $\delta$ is denoted as 1 if $l_i \neq RA_i$, otherwise 0. Subsequently, this random sampling procedure was repeated 1000 times, resulting in 1000 instability values for each category number. By comparing the instability values across different category numbers, we found that the instability values for category numbers 2, 3 and 7 were significantly lower than those of other category numbers (nonparametric permutation tests, 20000 times, $P < 0.001$, Bonferroni corrected, Fig. S1A), suggesting the clustering in these three situations was stable across populations.

Second, in the null model analysis, we tested if the spatial pattern among the classified categories was significantly different from the random situation. Briefly, for each individual, we first estimated the metric-to-metric spatial similarity matrix and obtained the classification results under categories 2 to 7 by using HCA, and these classifications were referred to as the individual assignment. Then, 100 corresponding random networks with same size and degree distribution were generated for each individual and the metric classifications were similarly estimated as the random assignments. Thus, for each category number, the disagreement in the classification results between the individual assignment and each of the random assignments could be assessed by using the abovementioned instability value formula and these instability values were further averaged for each individual to represent the differences in individual brain networks and the corresponding random situations. Finally, for each category number, we examined whether its averaged instability value was significantly higher than zero across individuals, and we found that except for the use of 2 categories, the brain networks were significantly different from random situations (nonparametric permutation paired tests, 20000 times, $P < 0.001$, Bonferroni corrected, Fig. S1B). Collectively, both 3 and 7 categories corresponded to across-population stability and topological uniqueness of brain networks. Considering that 7 categories are less meaningful, we classified the nodal metrics into 3 categories for further analyses.

**R-fMRI Data Preprocessing**

The routine preprocessing of R-fMRI data was performed using DPABI [7] for each participant. In detail, the first 5 volumes were removed, and the remaining volumes were corrected for slice timing and head motion. Then, the T1-weighted image was co-registered to the mean functional image and was subsequently segmented into gray matter (GM), white matter (WM) and cerebrospinal fluid (CSF) by using a unified segmentation algorithm. The resultant GM, WM and CSF images were further nonlinearly registered into the Montreal Neurological Institute (MNI) space with the transformation parameters estimated. Then, the functional data were normalized to the MNI space by using the estimated transformation parameters and resampled to 3-mm isotropic voxels. Next, spatial smoothing was applied to the normalized functional images with a 4-mm full width half maximum (FWHM) Gaussian kernel. Then, the linear drift was detrended, and several nuisance signals, including the Friston's 24 head motion parameters, the signals from the whole brain, WM and CSF were regressed out to reduce respiratory and cardiac effects. Finally, temporal filtering (0.01 – 0.1 Hz) was performed on the time series of each voxel to reduce the effect of low-frequency drifts and high-frequency physiological noise.

**R-fMRI Brain Network Analyses**

*Construction of Functional Brain Networks.* A group-level voxel-wise functional network (5% network density) was constructed based on the preprocessed R-fMRI data. Briefly, for each individual, we extracted the time series of each voxel within a GM mask, which was defined by thresholding the priori GM probabilistic template in SPM8 (GM probabilistic density > 0.2). Then, the Pearson's correlation of each pair of the time series was estimated, resulting in a functional connectivity matrix for each individual. Finally, we averaged the connectivity matrices across individuals to generate a grouped-averaged matrix and the top 5% strong connections were selected to define the group-level weighted functional network.

*Identification of Functional Brain Systems.* To examine the specificity of the distributions of structural hubs in functional systems, we identify the functional systems based on our group-level voxel-wise functional network. Briefly, we first applied a spectral community algorithm [8] to the functional network and 13 functional modules were identified (modules with a size less than 100 voxels were removed). Then, we merged some of the 13 modules that were obviously sub-sets belonging to a large functional system, according to prior functional connectome studies [6]. Finally, we obtained a seven-system parcellation that included the visual, sensorimotor, dorsal attention, ventral attention, limbic, frontoparietal and default-mode systems (Fig. S2A).

*Functional Participant Coefficients.* The functional participant coefficient quantifies the level that a given node connects to different functional systems [9, 10, 11]. We calculated the functional participant coefficient (FPc) for each node (i.e., voxel) in our group-level voxel-wise functional network by using the following formula,

$$FPc_i = 1 - \sum_{s=1}^{S}(Dc_{is}/Dc_i)^2$$

where, $S$ is the number of functional modules, $Dc_{is}$ specifies the degree of node $i$ within module $s$, and $Dc_i$ specifies the degree centrality of node $i$ (Fig. S2B).

**Validation Analysis**

To determine whether our findings are robust for use under different diffusion imaging protocols, reconstructing algorithms of fiber pathways, node definitions, and hub selection thresholds, we implemented the validation analyses via three procedures: *i) The Effects of the Diffusion Imaging Protocol and the Fiber Reconstructing Algorithm.* It has been argued that DTI tractography approaches can introduce false negative long-range connections and false positives in tracing between nearby regions because of their inaccuracies in resolving crossing fibers and tracts with sharp angles [12]. Thus, to determine whether our main findings are insensitive to the diffusion imaging protocol and whether they are influenced by the fiber pathway reconstruction algorithm, we utilized the HARDI data from Dataset 3 to reconstruct individual structural brain networks with 1024 nodes. Specifically, we obtained the minimal preprocessing HARDI data with eddy current and susceptibility distortion correction from the HCP website (http://db.humanconnectome.org) [2]. The reconstruction of the diffusion profile was then implemented in a voxel-by-voxel manner using a generalized q-sampling imaging model [13]. Furthermore, whole-brain fiber tracts were generated, and individual structural networks were constructed. Finally, we identified the hub categories and examined their spatial distribution and miscellaneous characteristics. *ii) The Effects of Node Definition of the Structural Network.* Previous studies have demonstrated that different node definitions used during brain network construction can lead to differences in network topological properties [14, 15]. Using Dataset 1, we validated whether our main findings were affected by another regional parcellation with 625 nodes that were generated based on the constraint of the anatomical transcendental boundaries of automated anatomical labeling. Network construction and analyses were performed again, as described previously. *iii) The Effects of Hub Selection Thresholds.* In this study, the brain nodes with the top 20% of hub indices were defined as hubs, which could influence our conclusions. Therefore, based on Dataset 1, we also selected two additional thresholds, the top 15% and the top 25% of hub indices, to define brain hubs. The hub characteristics were explored again to verify our main findings.

## Supplemental Results

### Validation Results

*Data using Different Imaging and Tractography Protocols.* We validated the main findings by re-performing our analysis on Dataset 3 (HARDI data from HCP). We found remarkably similar or different spatial distributions between specific network nodal metrics (range of Spearman's ρ: 0.29~1.00), which were highly similar to the main results (Fig. S4A and 3A). The HCA classified the eight-nodal metric maps into three categories: i) eigenvector centrality, subgraph centrality, K-core centrality, closeness centrality, and degree centrality; ii) betweenness centrality and page-rank centrality; and ii) participant coefficient (Fig. S4A). Of note, degree centrality was classified into the first category, which might suggest the bipolar topological character of degree centrality and/or its potential sensitivity to different diffusion imaging protocols or fiber reconstructing algorithms. Although the classification result was slightly changed, the spatial distributions of all three hub indices over the whole brain and within functional systems were almost the same as the main findings, and the commonly identified hub nodes in all three categories were primarily located at the default-mode system ($Ps < 0.005$, Bonferroni corrected) with additionally identified hub nodes in the visual system for aggregated hubs and in the sensorimotor and ventral attention systems for connector-hubs ($Ps < 0.01$, Bonferroni corrected) (Fig. S4B and S4C). Moreover, all three categories of hubs exhibited better microstructural organization, greater wiring costs, higher functional association, more cognitive flexibility and heavier topological vulnerability than non-hubs. Among the three categories of hubs, the aggregated hubs exhibited the largest generalized FA values and the longest streamline length, the distributed hubs exhibited the highest streamline cost and topological vulnerability, and the connector hubs exhibited strongest functional association and highest cognitive flexibility (Fig. S5). These results indicate the strong reproducibility of our findings under different imaging and tractography methods.

*Different Node Definitions.* We used a 625-node definition to re-construct the whole-brain WM individual networks and found that the results were largely consistent with our results in the main text. Briefly, the classification of the three categories of metrics was exactly the same as the classification used in the main text, and the spatial patterns and system distribution of hub indices were largely consistent with the main results (Fig. S6). Moreover, the distinct miscellaneous characteristics of different structural brain hubs were extremely retained, which were highly similar to those of the networks with 1024 nodes (Fig. S7). Collectively, our main findings were independent of the node definition during structural brain network construction.

*Thresholds for Hub Identification.* The nodes with the top 20% of hub indices were identified as the hubs in the main analysis, and two additional thresholds, 15% and 25%, were used for validation analyses. We found that either under the threshold of 15% (Fig. S8) or 25% (Fig. S9), all three categories of hubs exhibited better microstructural organization, greater wiring costs, higher functional association, more cognitive flexibility and heavier topological vulnerability than non-hubs, and the diversity of these characteristics among brain hubs was highly consistent with the main findings.

# Supplemental Tables

**Table S1.** Detailed description of eight graph nodal metrics

| Nodal Metric Name (Abbreviation) | Description | Formula | Annotation |
|---|---|---|---|
| Betweenness Centrality (Bc) | Freeman's betweenness centrality specifies the number of times that a node is on the shortest path between two other nodes in network [16]. | $Bc_k = \sum_{i=1}^{N} \sum_{j=1}^{N} g_{ikj}/g_{ij}$ | $N$ is the number of nodes, $g_{ikj}$ is the number of shortest paths between node $i$ and node $j$ passing through node $k$, and $g_{ij}$ is the total number of all shortest paths between node $i$ and node $j$. |
| Closeness Centrality (Cc) | Freeman's normalized closeness centrality is defined as the reciprocal of averaged distance across all shortest paths between a given node and the other nodes [17]. | $Cc_i = (N-1)/\sum_{j=1}^{N}(d_{ij})$ | $N$ is the number of nodes, and $d_{ij}$ is the length of shortest path between node $i$ and node $j$. |
| Degree Centrality (Dc) | Degree centrality is calculated as the number of edges connected to a given node [17]. | $Dc_i = \sum_{j=1}^{N} A_{ij}$ | $N$ is the number of nodes, and $A$ is the adjacent matrix, if there is an edge between node $i$ and node $j$, $A_{ij} = 1$, otherwise $A_{ij} = 0$. |
| Eigenvector Centrality (Ec) | Eigenvector centrality is the principal eigenvector of the adjacency matrix [18]. In particular, it mathematically equivalent to Katz's centrality [19] as the damping factor approaches the reciprocal of the principal eigenvalue from below [20], and it is the weighted count of all walks for a given node that considers indirect paths. | $Ec_i = \mu_i^1 \sim \sum_{j=1}^{N} \sum_{k=1}^{+\infty} (1/\lambda_1)^k (A^k)_{ij}$ | $N$ is the number of nodes, $\mu_i^1$ is the $ith$ component of the principal eigenvector, $\lambda_1$ is the largest eigenvalue of the adjacency matrix, $A$ is the adjacent matrix, and $(A^k)_{ij}$ specifies the path between node $i$ and node $j$ with $k$ step walking. |
| K-core Centrality (Kc) | K-core decomposition assigns a set of nodes to k if and only if the minimum degree of the subgraph comprised of these nodes is k [21]. It assesses the level of interconnection between each other for a given set of nodes [9]. | $G = (V, E),$ $H = (C, E\|C), C \subseteq V,$ $\forall v \in C, Dc_v \geq k,$ $Kc_C = k$ | $G$ represents a graph, $V$ is the node set, $E$ is the edge set, $H$ is the subgraph, $v$ is a given node in subgraph $H$, $Dc_v$ specifies the degree centrality of node $v$. |
| Participant Coefficient (Pc) | Participant coefficient quantifies the level that a given node connects to different network modules [9, 10, 11]. Modularity detection utilized a spectral community algorithm [8]. | $Pc_i = 1 - \sum_{s=1}^{S}(Dc_{is}/Dc_i)^2$ | Where, $S$ is the number of modules, $Dc_{is}$ specifies the number of edges between node $i$ and the other nodes within module $s$, and $Dc_i$ specifies the degree centrality of node $i$. |
| Page-rank Centrality (Pr) | Google's page-rank centrality [22] is a variant of the eigenvector centrality [23]. The damping factor was set to 0.85, which was generally used in previous studies and introduced a small probability walking on the graph [24]. | $Pr_i = (1-d)/N + d\sum_{j=1}^{N}(A_{ij}/Dc_i)$ | $N$ is the number of nodes, $d$ is the damping factor, $A$ is the adjacent matrix, and $Dc_i$ specifies the degree centrality of node $i$. |
| Subgraph Centrality (Sc) | Subgraph centrality [25] quantifies the number of subgraphs in which a given node is included. | $Sc_i = \sum_{k=0}^{+\infty}(A^k)_{ii}/k! = \sum_{j=1}^{N}\mu_{ij}^2 e^{\lambda_j}$ | $(A^k)_{ii}$ is the number of subgraphs with $k$ step walking, $\mu_{ij}$ is the $ith$ component of $jth$ eigenvector, and $\lambda_j$ specifies $jth$ eigenvalue of the adjacent matrix $A$. |

# Supplemental Figures

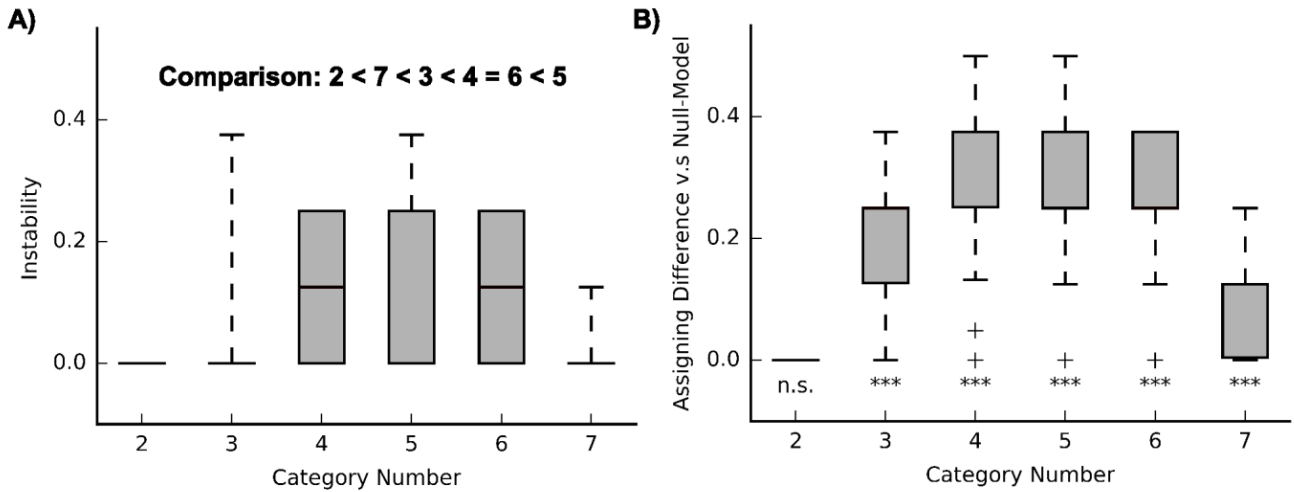

**Figure S1. Identification of the Category Number of the Hierarchical Clustering Analysis.** (A) The instability of classification with different category numbers. The symbol "<" indicates that the item to the left is significantly lower than the item to the right ($P < 0.001$, Bonferroni corrected), and the symbol "=" represents no significant differences between the left and right items. (B) The assigning differences compared with null models when using different category numbers. The symbol "***" indicates that the assigning difference is significantly larger than 0 ($P < 0.001$, Bonferroni corrected), and the symbol "n.s." indicates not significant.

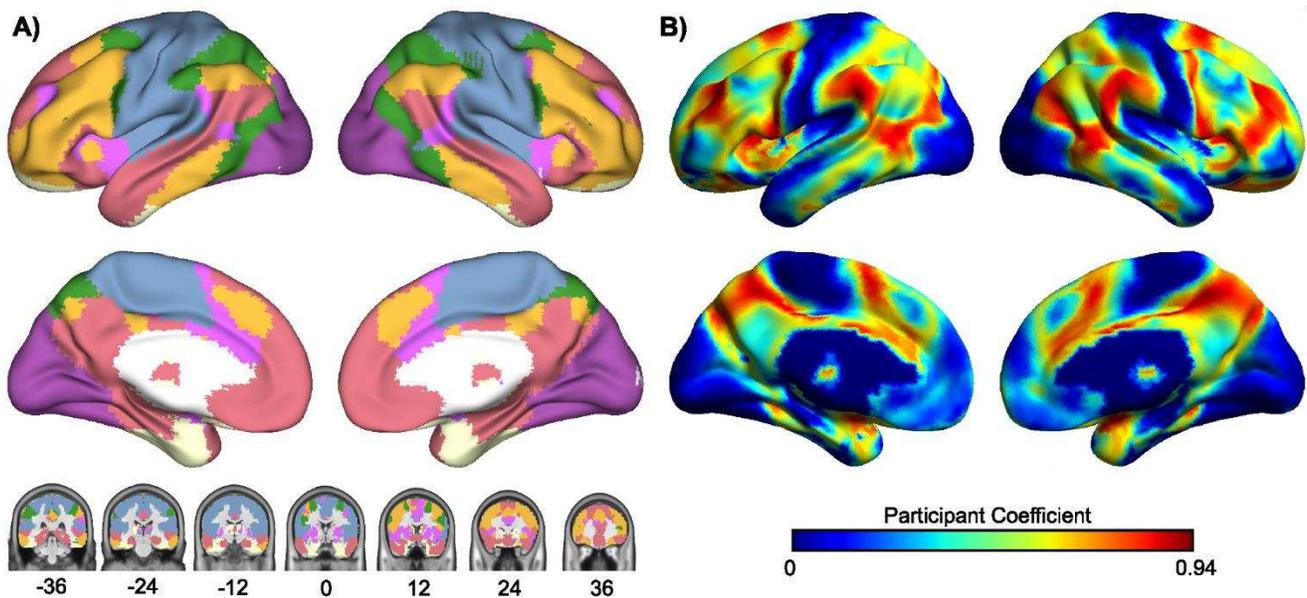

**Figure S2. Brain Functional Systems and Functional Participant Coefficient.** (A) Seven functional systems identified by using the fMRI data in Dataset 1. These functional systems were mapped to a brain surface and also to 8 coronal slices (subcortical regions: MNI coordinates from y = -36 to 36 with steps of 12 mm). (B) The corresponding functional participant coefficient distribution of the group-level voxel-wise functional network.

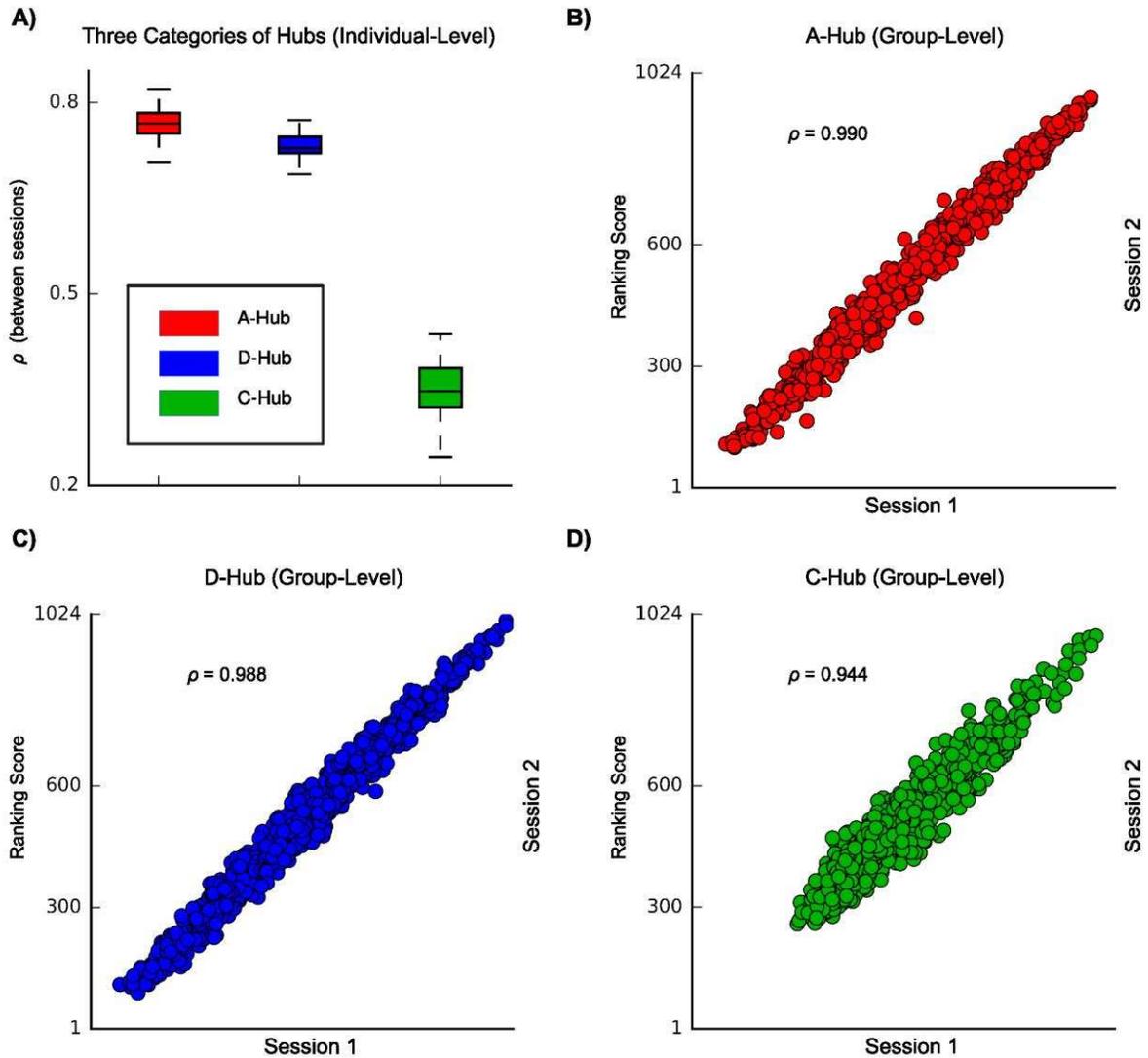

**Figure S3. Spatial Correlation between the Three Categories of Hub Indices Obtained from Dataset 2.** (A) Individual-level spatial correlation. For each box plot, the bottoms and tops of the boxes indicate the first and third quartiles of the Spearman's correlation coefficients across individuals, the band inside the box represents the median, and the whiskers specify the 1.5 interquartile range of the lower and upper quartiles. (B, C, D) Group-level spatial correlation. The red, blue and green plots indicate the aggregated, distributed and connector hub indices, respectively.

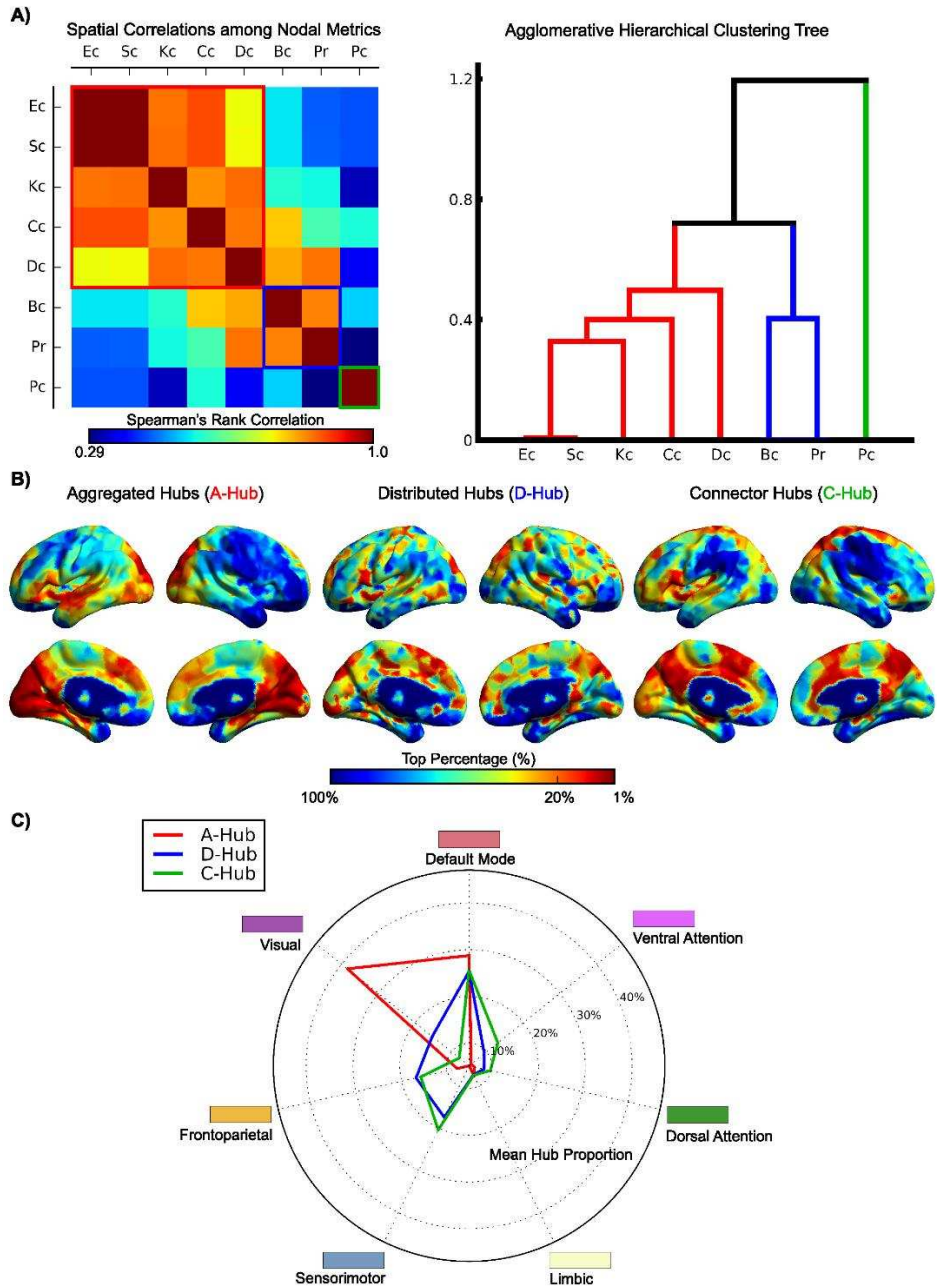

**Figure S4. Identification and Spatial Distribution of the Three Categories of Hubs from Dataset 3.** (A) The group-averaged map of the Spearman's correlations among the eight nodal metrics and the agglomerative hierarchical clustering tree generated from the map. The red, blue and green solid lines show the results of the classification, indicating the three categories of metrics used to identify the following aggregated hubs, distributed hubs and connector hubs. (B) Spatial distributions of the three categories of hubs on the brain surface. (C) Spatial distributions of three categories of hubs in the seven functional systems.

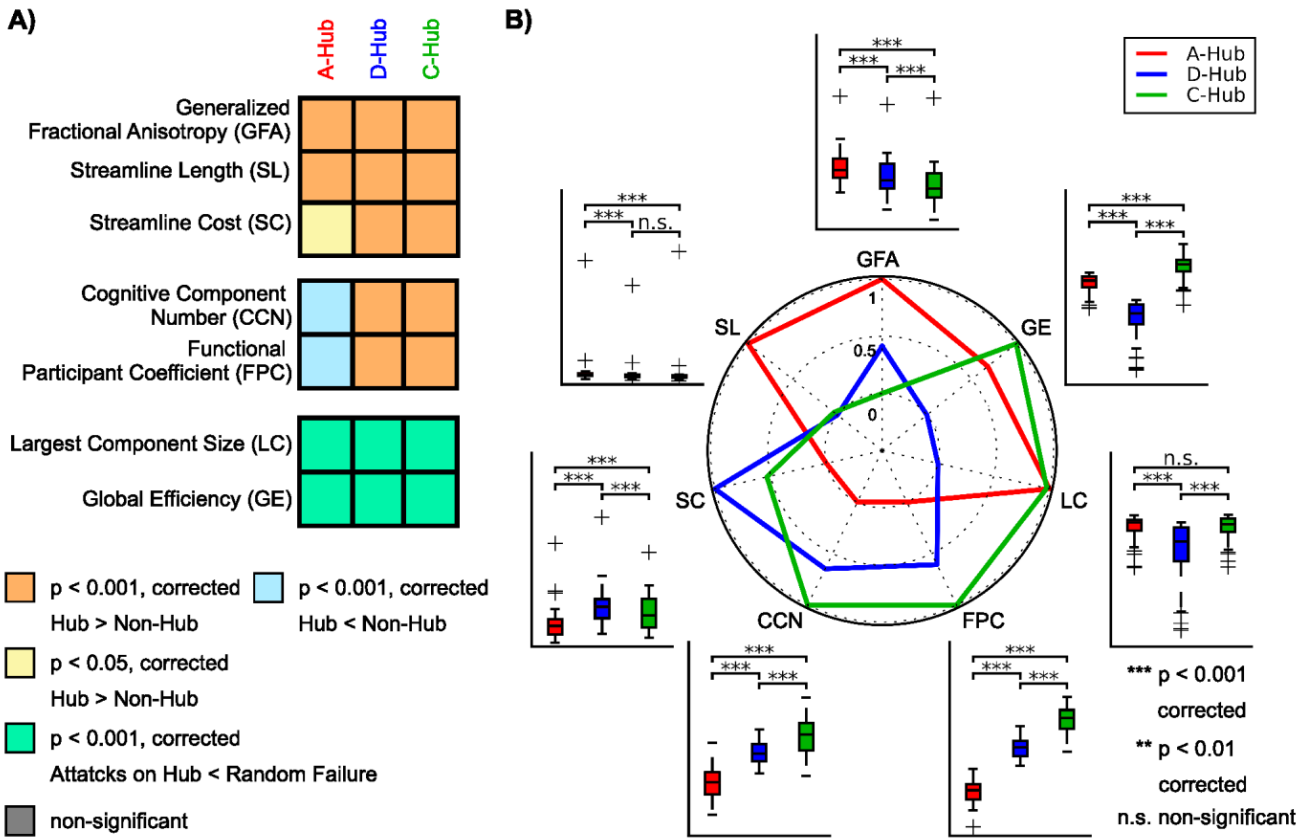

**Figure S5. The Miscellaneous Characteristics of the Three Categories of Hubs from Dataset 3.** (A) Comparisons of miscellaneous characteristics between hubs and non-hubs for each category of hubs. (B) Comparisons of these characteristics among the three categories of hubs.

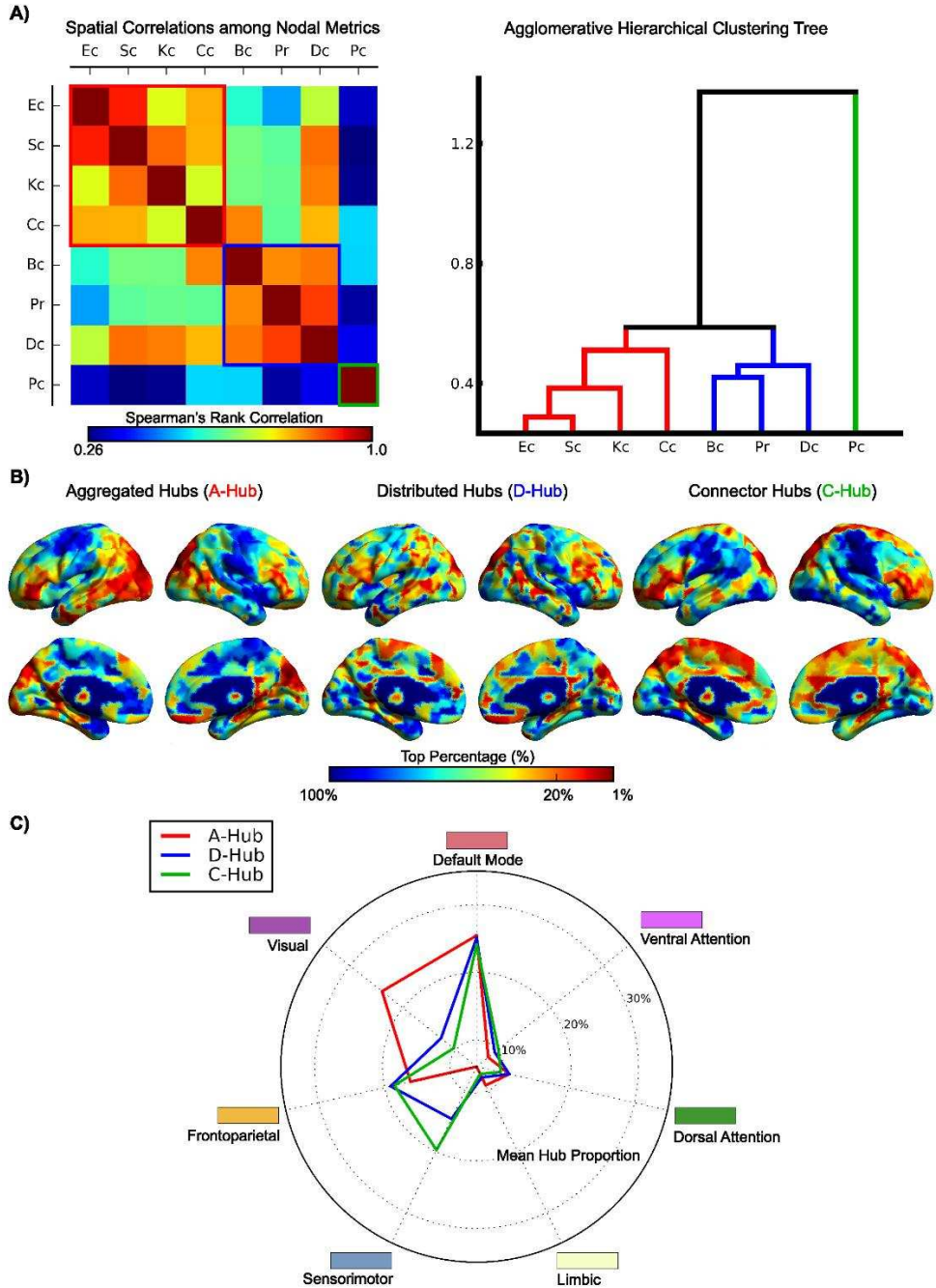

**Figure S6. Identification and Spatial Distribution of the Three Categories of Hubs using the 625-Node Definition.** (A) The group-averaged map of Spearman's correlations among eight nodal metrics and the agglomerative hierarchical clustering tree generated from the map. The red, blue and green solid lines show the results of the classification, indicating the three categories of metrics used to identify the following aggregated hubs, distributed hubs and connector hubs. (B) Spatial distributions of the three categories of hubs on the brain surface. (C) Spatial distributions of the three categories of hubs in the seven functional systems.

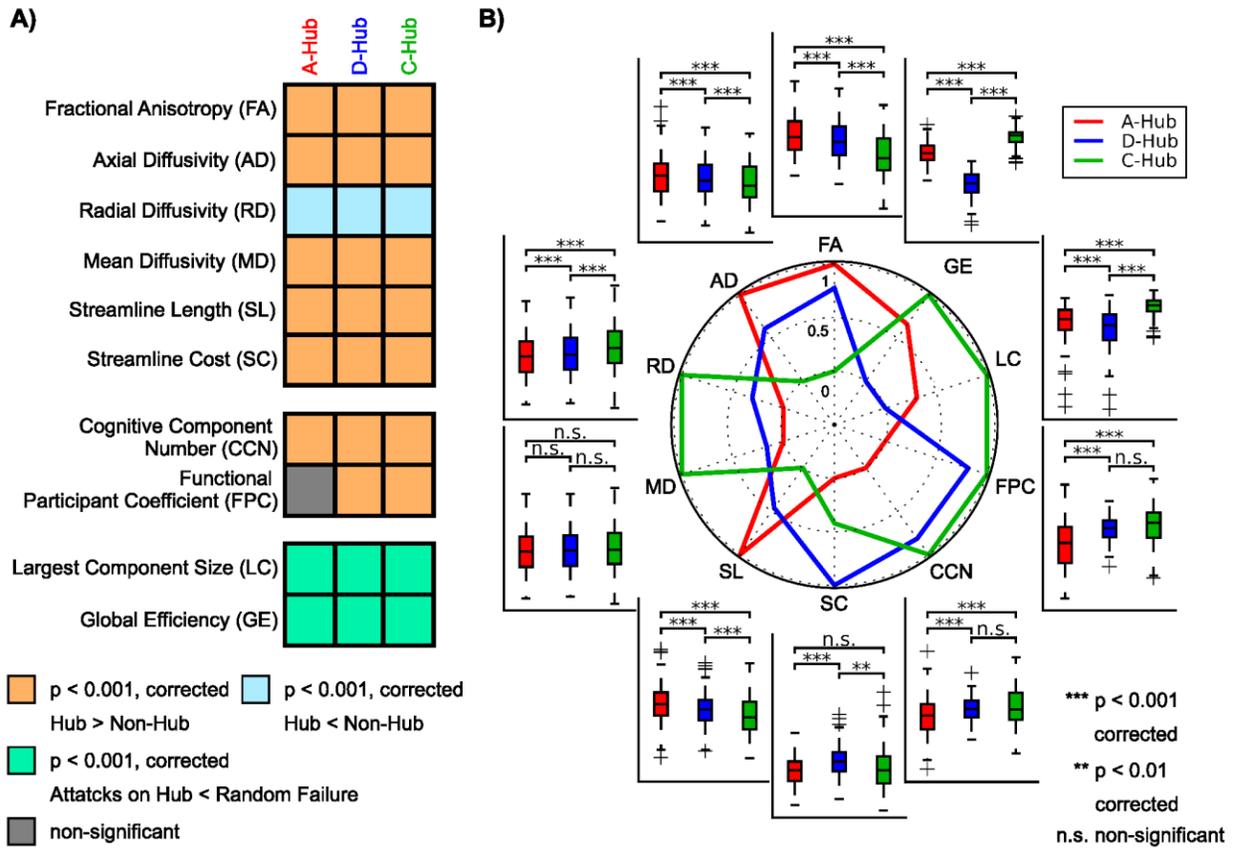

**Figure S7. The Miscellaneous Characteristics of the Three Categories of Hubs using the 625-Node Definition.** (A) Comparisons of the miscellaneous characteristics between hubs and non-hubs for each category of hubs. (B) Comparisons of these characteristics among the three categories of hubs.

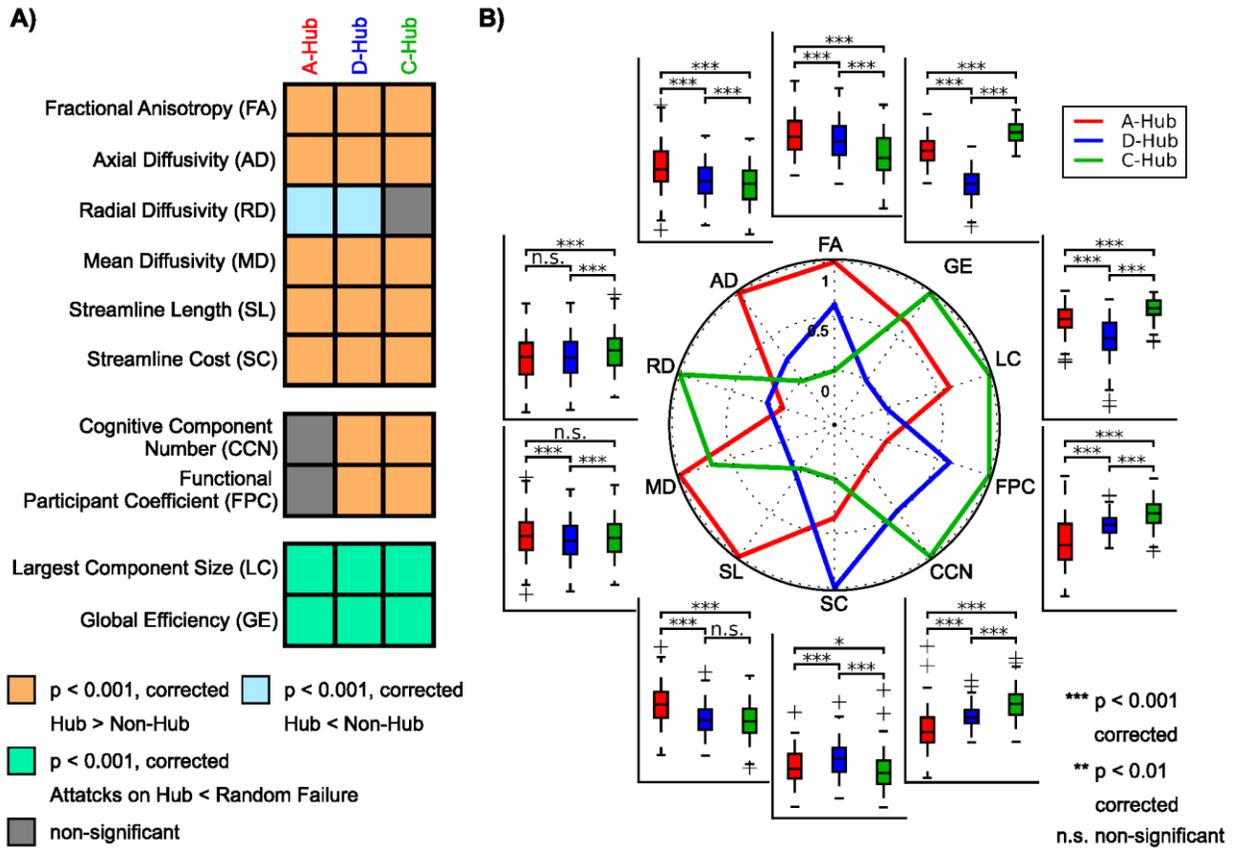

**Figure S8. The Miscellaneous Characteristics of the Three Categories of Hubs using the 15% Hub Selective Threshold.** (A) Comparisons of the miscellaneous characteristics between hubs and non-hubs for each category of hubs. (B) Comparisons of these characteristics among the three categories of hubs.

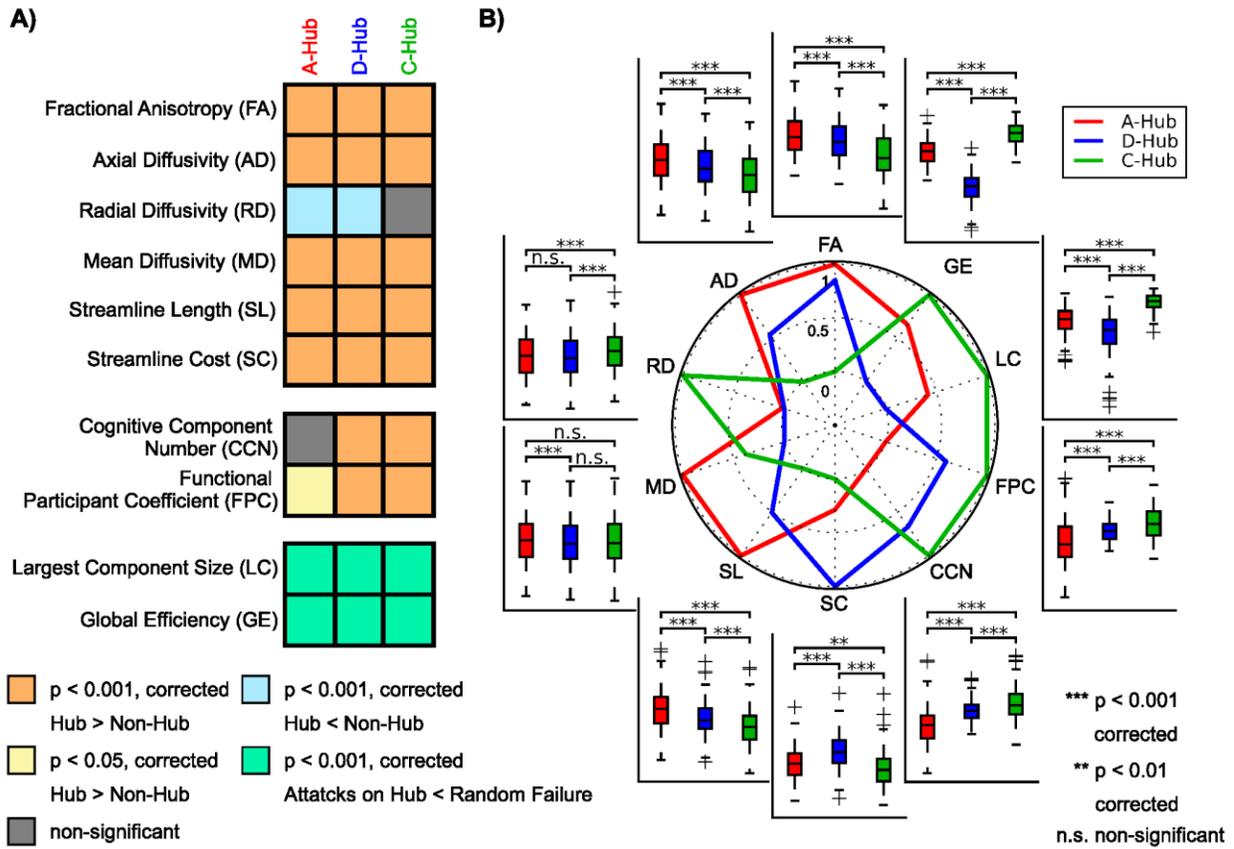

**Figure S9. The Miscellaneous Characteristics of the Three Categories of Hubs using the 25% Hub Selective Threshold.** (A) Comparisons of the miscellaneous characteristics between hubs and non-hubs for each category of hubs. (B) Comparisons of these characteristics among the three categories of hubs.